\documentclass[12pt,twoside,a4paper]{article}
\usepackage{graphicx,lipsum,float,enumerate,caption,framed,emptypage}

\usepackage{amsmath}
\usepackage{amsfonts}
\usepackage{amssymb}
\usepackage[scale=0.80]{geometry}
\title{\textbf{Dynamics of quantum observables and Born's rule in Bohmian Quantum Mechanics}}

\author{Athanasios C. Tzemos\footnote{atzemos@academyofathens.gr} \, and George Contopoulos\\ Research Center for Astronomy and Applied Mathematics \\of the  Academy of Athens\\  Soranou Efessiou 4, GR-11527 Athens, Greece }

\date{}

\begin{document}
\maketitle
\begin{center}\textbf{Abstract}\end{center}

We investigate both ordered and chaotic Bohmian trajectories within the Born distribution of Bohmian particles of an anisotropic 2d quantum harmonic oscillator. We compute the average values of energy, momentum, angular momentum, and position using both Standard Quantum Mechanics and Bohmian Mechanics. In particular, we examine realizations of the Born distribution for a wavefunction with a single nodal point and two different wavefunctions with multiple nodal points: one with an almost equal number of ordered and chaotic trajectories, and another composed primarily of chaotic trajectories. Throughout our analysis, we focus on elucidating the contribution of ordered and chaotic Bohmian trajectories in determining these average values.



\section{Introduction}
Bohmian Quantum Mechanics (BQM)  is a basic alternative interpretation of Quantum Mechanics \cite{Bohm,BohmII,holland1995quantum,bacciagaluppi2009quantum, lazarovici2019quantum}.  In BQM the quantum particles follow deterministic trajectories, which are guided by the wavefunction describing the system, i.e. the solution of the Schr\"{o}dinger equation:
\begin{equation}
-\frac{\hbar^2}{2M_q}\nabla^2\Psi+V\Psi=i\hbar\frac{\partial{\Psi}}{\partial t},\quad q=x,y,\dots
\end{equation}
according to the so called ``Bohmian equations of motion'':
\begin{equation}
M_q\frac{dr}{dt}=\hbar\Im\left(\frac{\nabla \Psi}{\Psi}\right).
\end{equation}
Thus the state of a quantum system in BQM is described by its wavefunction and the positions of the associated Bohmian particles in the configuration space.

BQM predicts the same experimental results as Standard Quantum Mechanics (SQM), when the Bohmian particles are distributed according to Born's rule (BR), which states that the probability density of finding a quantum particle in a certain region of space is equal to $P=|\Psi|^2$. If BR is initially satisfied then it is easily proved that it will be satisfied for all times. However, in BQM we are allowed to consider initial configurations of Bohmian particles distributed with $P_0\neq |\Psi_0|^2$. Whether BR will be accesible in the long run in such cases is a fundamental problem in BQM and  it has been shown that it depends strongly on the complexity of the Bohmian trajectories \cite{valentini1991signalII,valentini1991signalI,towler2011time}.

Chaos in Quantum Mechanics is an open field of research in the last decades and has attracted the interest of many authors. In Classical Mechanics we have a well understood definition of chaos: A nonlinear dynamical system is chaotic when it has a bound phase space and exhibits high sensitivity on small changes of its initial conditions. However, Standard Quantum  Mechanics \cite{merzbacher1998quantum,shankar2012principles,ballentine2014quantum},
 does not predict trajectories for the quantum particles (due to the existence of Heisenberg's uncertainty). Moreover, the evolution of the wavefunction is described by Schr\"{o}dinger's equation, i.e. a linear equation. Thus the notion of quantum chaos is not clearly defined in SQM and the majority of works in the field are focused on the comparison between the properties of quantum systems whose classical counterparts are ordered or chaotic \cite{haake1991quantum, Stockmann_1999, wimberger2014nonlinear,robnik2016fundamental}.

In contrast with SQM, BQM is a highly nonlinear trajectory-based quantum theory which allows the coexistence of ordered and chaotic Bohmian trajectories for a given quantum system.  Thus it allows the study of chaotic dynamics in the quantum framework by use of all the techniques of the theory of classical dynamical systems, something that has been the main topic of many works \cite{parmenter1995deterministic,sengupta1996quantum,iacomelli1996regular,frisk1997properties, wu1999quantum, makowski2000chaotic, makowski2001simplest,makowski2002forced, wisniacki2003dynamics, falsaperla2003motion, wisniacki2005motion,  wisniacki2007vortex,borondo2009dynamical, cesa2016chaotic,santos2024broglie}.

In our series of works on Bohmian chaos we showed that  the nodal points of the wavefunction $\Psi$ (i.e. the points where $\Psi=0$) are essential for the production of chaos. In fact, in the frame of reference of a moving nodal point $N$ there is an unstable fixed point, the `X-point'. Together they form the so called `nodal point-X-point complexes' (NPXPC) which are characteristic geometrical structures of the Bohmian flow in the close neighbourhood of a nodal point. When a Bohmian particle comes close to an NPXPC it gets deflected by the X-point \cite{efth2009}. The effect of many such encounters is the emergence of chaos. Furthermore, in a  recent paper \cite{tzemos2023unstable} we showed that there are also unstable points in the inertial frame of reference, the `Y-points' that play  some role in the generation of chaos. Trajectories that never approach X-points or  Y-points are in general ordered. 

The study of the mechanism responsible for chaos production was crucial for understanding the role of order and chaos in the accessibility of BR by arbitrary initial distributions of Bohmian particles. In \cite{tzemos2020chaos,tzemos2021role} we emphasized the fact that any realization of the  BR distribution made out of a finite number of Bohmian particles contains initial conditions that lead, in general, partly to chaotic and partly to  ordered trajectories.  Moreover, the chaotic trajectories of the systems that we studied  were found to have a common footprint on the configuration space, i.e. their points acquire a common long time distribution regardless of their initial condition. We called this feature `ergodicity' (this definition is  different from the standard notion of ergodicity based  on the metric transitivity in phase space). Thus the chaotic trajectories  are ergodic (or partially ergodic, when the chaotic trajectories are divided into sets with different long time distributions \cite{tzemos2023unstable}). 

On the the other hand, the corresponding long time distributions of points of ordered trajectories cover limited regions of the configuration space. Thus we found that BR will be accessible only by  initial distributions  with the correct proportion between ordered and chaotic trajectories, and  the correct distribution of ordered trajectories. It is also remarkable that  the number of ordered trajectories in the BR distribution may not be negligible even in the case of systems with multiple nodal points. Consequently, the common belief that a large number of nodal points guarantees the accessibility of Born's rule is in general not valid.

After all the above studies in the dynamics of Bohmian trajectories, it  is reasonable  for one to ask what is the role of the ordered and chaotic trajectories in deriving the same average values of observables as SQM, when $P_0=|\Psi_0|^2$. This is the subject of the present paper, where we consider a number of 2d wavefunctions and we calculate  the average values of the energy $E$, the momentum $(p_x,p_y)$,  the angular momentum $L$ and the position $(x,y)$ of a large number of Bohmian particles, sampled according to the Born distribution. These values are compared with the corresponding values predicted by Standard Quantum Mechanics.  By counting the numbers of ordered and chaotic trajectories in the Born distribution of these wavefunctions we examine their contribution in  the average values.

Most works in the field of Bohmian chaos concentrate on non-interacting bound systems such as harmonic oscillators, with only a few addressing interacting systems like the quantum Henon-Heiles system \cite{sengupta1996quantum, efthymiopoulos2006chaos}. In the present paper, we consider again a simple quantum system which is classically integrable, that of an anisotropic 2d harmonic oscillator $V=\frac{1}{2}(\omega_x^2x^2+\omega_y^2y^2$). In this case the solutions of the Schr\"{o}dinger  equation  are known analytically and the values of the various quantities can be calculated analytically as well. Details of these calculations are given in the Appendix. Moreover, the energy eigenstates of this system form a complete basis of Hilbert's space, i.e. any 2d wavefunction can be written as a linear combination of them. Consequently our results are not limited qualitatively in this special system.

We consider various wavefunctions which are   superpositions of 3 energy eigenstates of the 2d oscillator. In particular, we study a case with a single nodal point (Section 2) and two cases with many nodal points (Section 3), one  with equal proportions of  ordered and chaotic trajectories and one example where chaotic trajectories dominate the BR distribution. In Section 4 we make our summary and comment on our results.

\section{Wavefunction with a single node}

We consider first a wavefunction with a single nodal point:
\begin{equation}
\Psi=M(a\Psi_{0,0}+b\Psi_{1,0}+c\Psi_{1,1}).
\end{equation}
where $\Psi_{m,n}(x,y)=\Psi_m(x)\Psi_n(y)$. $\Psi_m(x)$ and $\Psi_n(y)$ are the 1-D energy eigenstates of the oscillator in $x$ and $y$ coordinates correspondingly, i.e.
\begin{equation}
\Psi_{m,n}=\prod_{q=x}^y N_q\exp\left(-\frac{\omega_qq^2}{2\hbar}\right)\exp\left(-\frac{i}{\hbar}E_{s}t\right)H_s\left(\sqrt{\frac{M_q\omega_q}{\hbar}}q\right),\label{mn}
\end{equation}
where $s=m,n$ (integers) for $q=x$ and $q=y$ respectively  and the normalization constant is  $N_q=\frac{1}{\sqrt{2^ss!}}\left(\frac{M_q\omega_q}{\pi\hbar}\right)^\frac{1}{4}$. The functions $H_m, H_n$ are Hermite polynomials in $\sqrt{\frac{M_x\omega_x}{\hbar}}x$ and   $\sqrt{\frac{M_y\omega_y}{\hbar}}y$ of degrees $m$ and $n$ respectively and $E_s=\left(s+\frac{1}{2}\right)\hbar \omega_q$.

\begin{figure}
\centering
\includegraphics[scale=0.3]{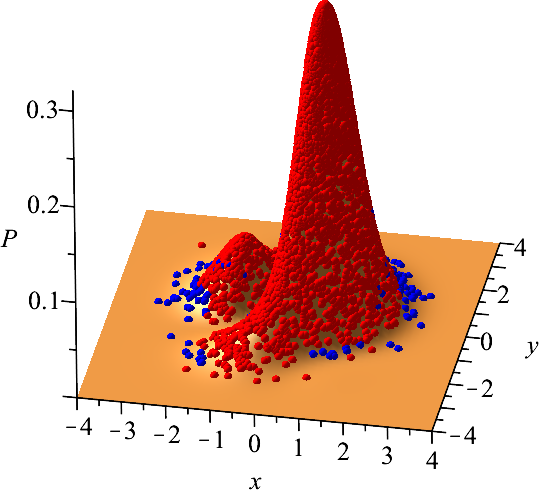}[a]
\includegraphics[scale=0.2]{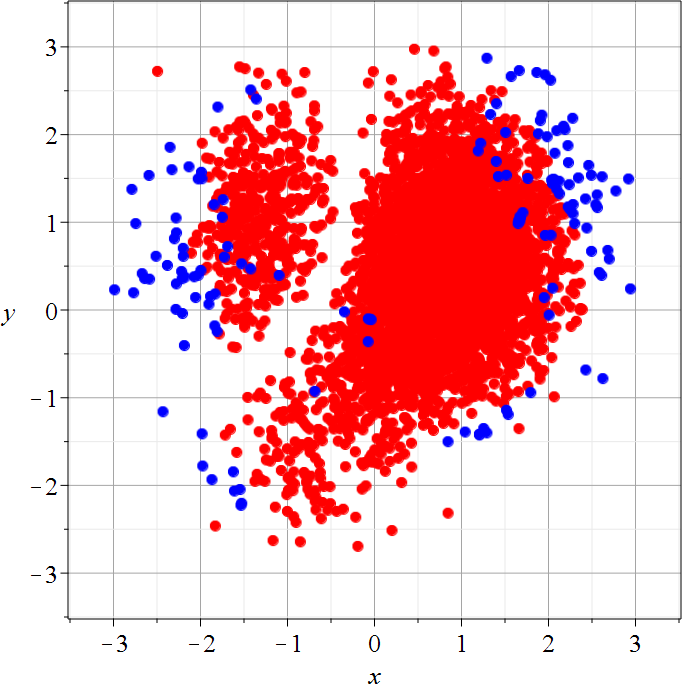}[b]
\includegraphics[scale=0.2]{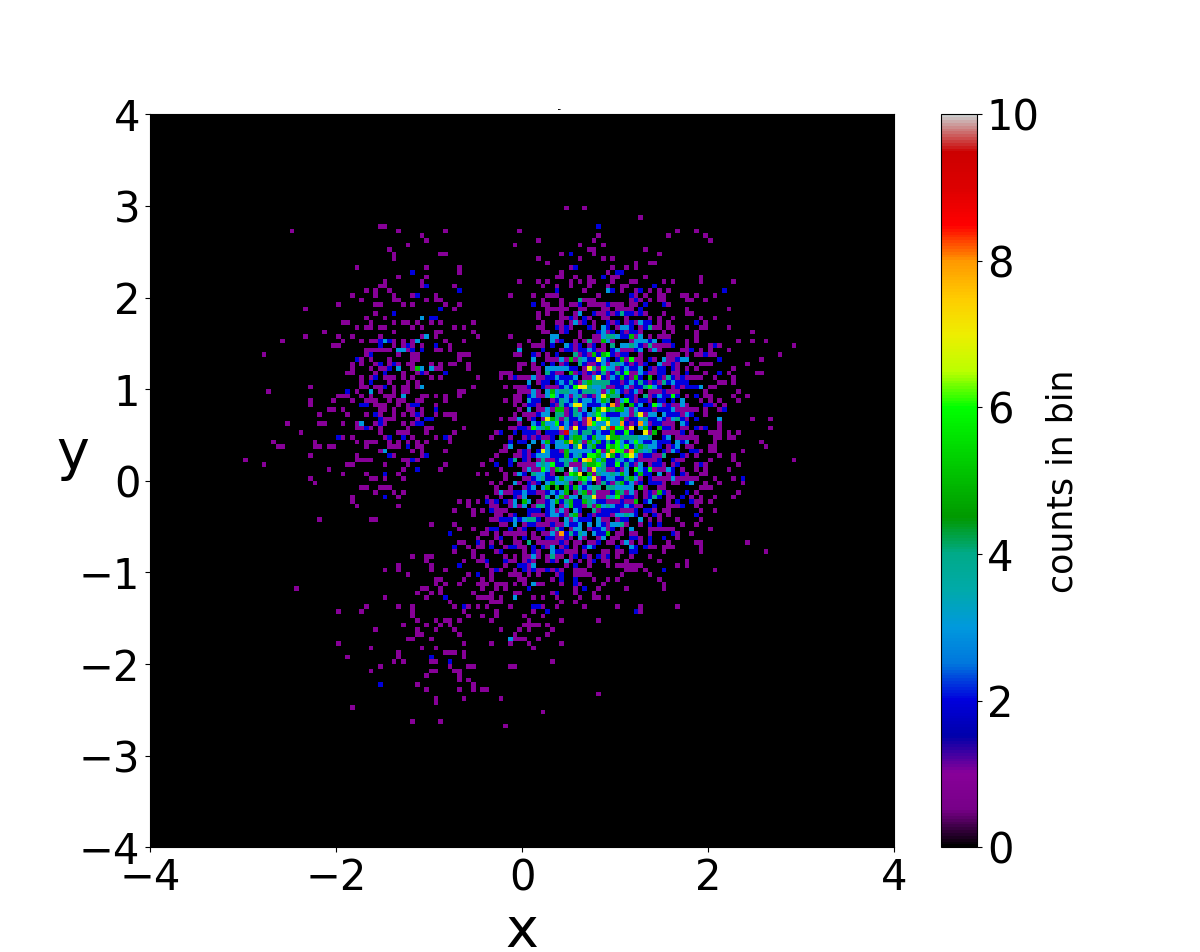}[c]
\includegraphics[scale=0.19]{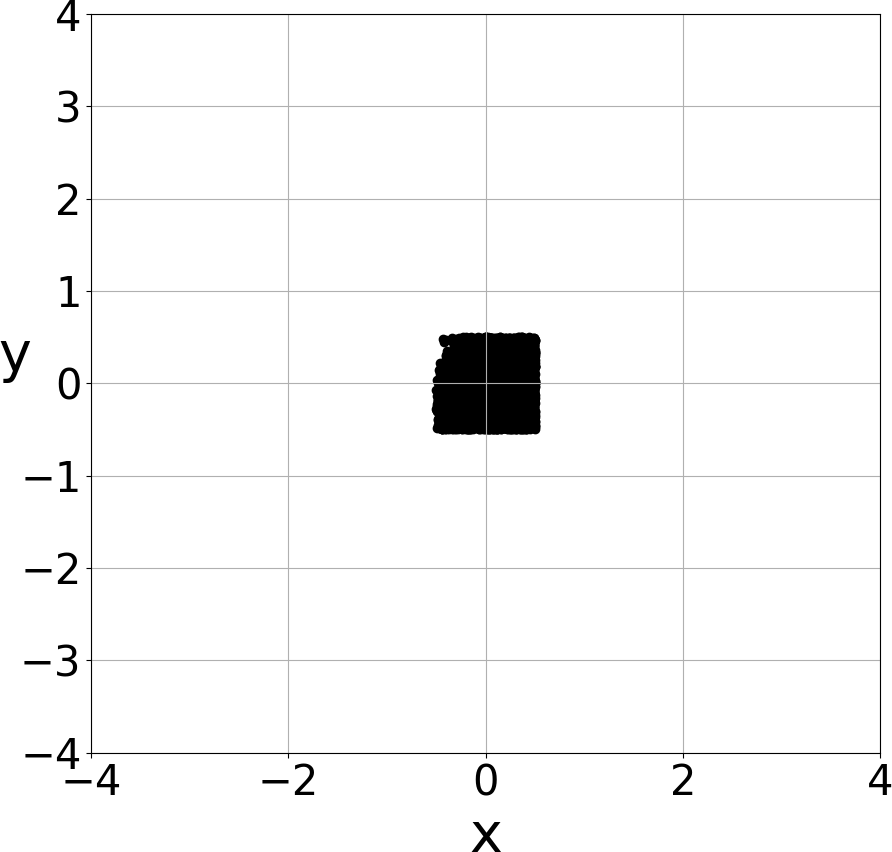}[d]
\caption{a,b,c: 5000 initial conditions of ordered (red) and chaotic (blue) trajectories distributed according to BR in the case of the wavefunction with a single nodal point: a) on the surface of $|\Psi|^2$  and b) on the $x-y$ plane. In c) we see their colorplot at $t=0$. d) 5000 non-Born distributed initial conditions.}\label{triplesin}
\end{figure}

In this case there is only one nodal point (the equations $\Psi_R=\Psi_{Im}=0$ are of first degree in $x$ and $y$).
In Fig.~\ref{triplesin} we show a realization of the Born distribution $P=|\Psi|^2$ made out of 5000 particles, when $M_x=M_y=1, \hbar=1, a=b=1, c=\sqrt{2}/2,\omega_x=1,\omega_y=\sqrt{2}/2$. The value of the normalizing constant  $M$ is found by setting $M^2(a^2+b^2+c^2)=1$. It is $M=\sqrt{2/5}$ \footnote{Although the wavefunction is defined from $-\infty$ to  $+\infty$ in $x$ and $y$ the probability density $P=|\Psi|^2$ takes appreciable values for all times in the central region of space around the origin. In the present paper we work in regions of space where where $P> 10^{-5}$ for all times.}. 

The red points represent the initial conditions which produce ordered trajectories and the blue points the initial conditions of chaotic trajectories.\footnote{We integrated for large times of order $10^5$ the 5000 initial conditions of the BR distribution and counted which of them cover a large part of the configuration space. The corresponding trajectories were characterized as chaotic. On the other hand the trajectories whose form was that of a deformed Lissajous figure covering a small region in space were characterized as ordered. This practical algorithm which is based on the long time form of the Bohmian trajectories, ordered and chaotic, was introduced and extensively used in our previous works (see for example  \cite{tzemos2023unstable,tzemos2021role}).} We observe that the ordered trajectories dominate the BR distribution (almost 97\%). The chaotic trajectories ($3\%$) start in regions of small $P=|\Psi|^2$ at the boundaries of the two blobs as shown in the plane $(x,y)$ (Fig.~\ref{triplesin}b).  

The maximum value of $P$ in the main blob is $P\simeq 0.33$ and in the secondary blob it is $P\simeq 0.06$. In Fig.~\ref{triplesin}c  we give with colors the density of the points of the trajectories at every $\Delta t=0.05$ of the BR distribution in a quadratic grid of bins up to a certain time $t$. Here we show the initial form of the colorplot at $t=0$. It is clear that the maxima correspond to the tops of the two blobs while $P$ decreases outwards.

Then we calculate the average value of the energy,  according to SQM, which is \cite{ballentine2014quantum}
\begin{align}\label{av}
\langle E\rangle=M^2(|a|^2E_{0,0}+|b|^2E_{1,0}+|c|^2E_{1,1}),
\end{align}
where 
\begin{equation}\label{ener}
E_{m,n}=(m+1/2)\hbar\omega_x+(n+1/2)\hbar\omega_y.
\end{equation}
We remind that since $\Psi_{m,n}$ are eigenstates of the quantum harmonic oscillator (i.e. stationary states)  $\langle E\rangle$ is constant in time. In the present case we find $\langle E\rangle=1.595$.

On the other hand, the average values of the energy $E_{av}$  of  the Bohmian particles  depend on their sampling in the realization of the Born distribution.  Each particle has a total energy 
\begin{equation}
E=K+V+Q,
\end{equation}
where $K$ is the kinetic energy $\frac{1}{2}(\dot{x}^2+\dot{y}^2)$, $V$ is the potential energy $V=\frac{1}{2}(\omega_x^2x^2+\omega_y^2y^2)$ and $Q$ is the Bohmian quantum potential energy $Q=-\frac{1}{2}\frac{\nabla^2|\Psi|}{|\Psi|}$ which is time dependent and thus $E$ is time dependent as well.

\begin{figure}
\centering
\includegraphics[scale=0.2]{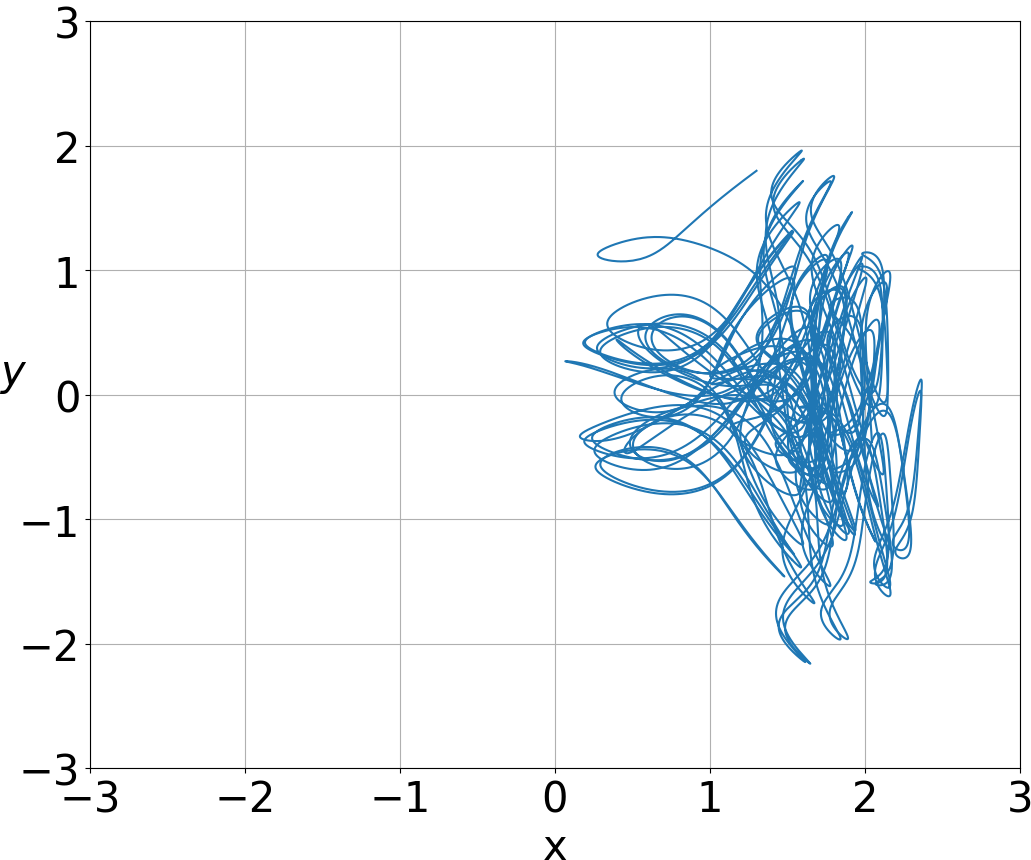}[a]
\includegraphics[scale=0.2]{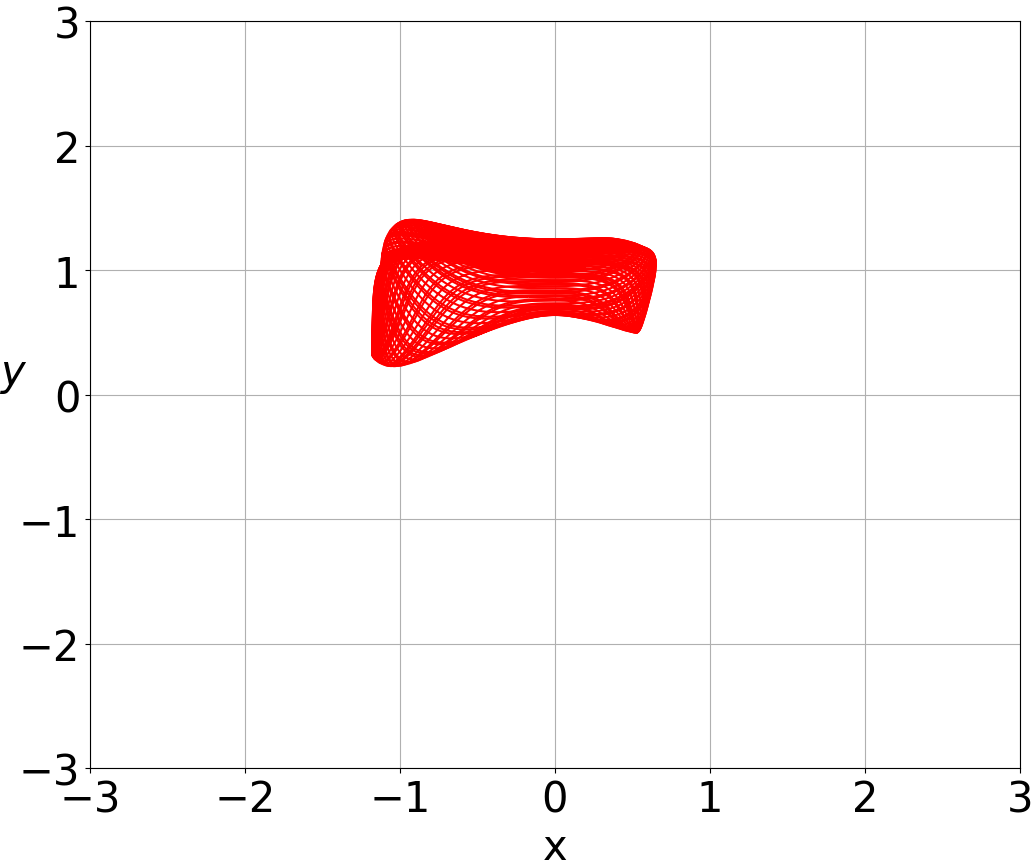}[b]
\caption{A chaotic trajectory ($x(0)=1.3 ,y(0)=1.8$) (a) and an ordered trajectory ($x(0)=0.2 ,y(0)=1.0$) (b) in the case of the single node wavefunction.}\label{chordsin}
\end{figure}

In Figs.~\ref{chordsin}a,b. we show typical examples of Bohmian trajectories, one chaotic and one ordered. We observe the complex shape of the chaotic trajectory and the close to Lissajous figure shape of the ordered trajectory, which covers a small region in the configuration space.

\begin{figure}
\centering
\includegraphics[scale=0.25]{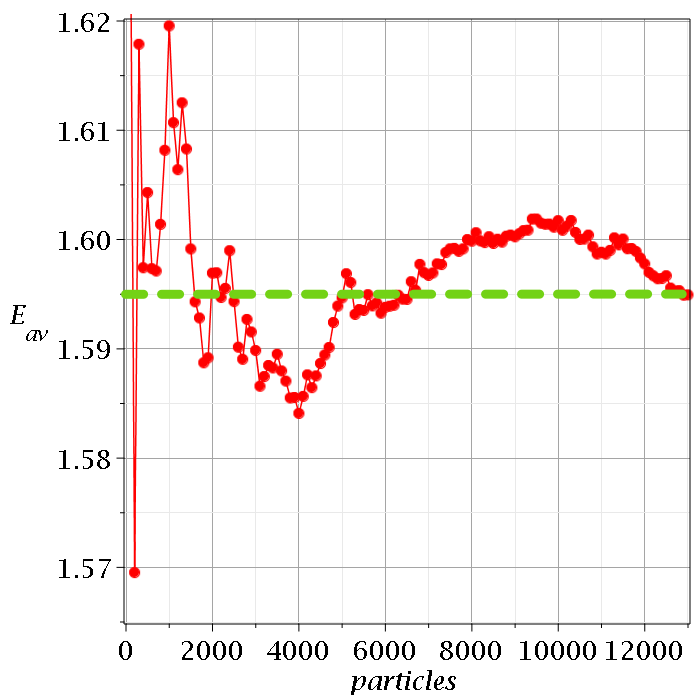}
\caption{The average value of energy $E_{av}$ at $t=0$ as a function of the number of particles in a realization of Born's distribution for the wavefunction with a single nodal point.}\label{avsin}
\end{figure}

In Fig.~\ref{avsin} we take the average value $E_{av}$ of the energy for various numbers of particles $N=100,200,\dots 13000$ of a realization of the BR distribution  at $t=0$. These values show  significant deviations from the analytical one for small numbers of Bohmian particles but for $N>5000$ {they tend to to approach it.}

For $N=5000$ the average value is close to $E_{av}=1.595$. But, in order to have a better estimation of $E_{av}$ we have considered 5 different (random) realizations of the BR distribution with $N=5000$ Bohmian particles. Their average value at $t=0$ is $E_{av}(0)=1.599\pm 0.004$, i.e. it is a little above $\langle E\rangle$. If we integrate the trajectories of all realizations up to $t=1000$ then we find that the mean absolute deviation between the analytical and the numerical values is $\sim 0.005$, i.e. we can write $E_{av}=\langle E\rangle\pm0.005$. Thus $E_{av}$ has deviation from $\langle E\rangle=1.595$ of order about 3\%. By taking larger numbers of particles we find a better approach to $\langle E\rangle$, as expected. 

On the other hand, in initial distributions of particles different from Born's rule we find quite different values of $E_{av}$. In a particular case with $P_0\neq |\Psi_0|^2$ (Fig.~\ref{triplesin}d) we found $E_{av}=0.819,$ a value much smaller than $\langle E\rangle=1.595$.

\begin{figure}
\centering
\includegraphics[scale=0.2]{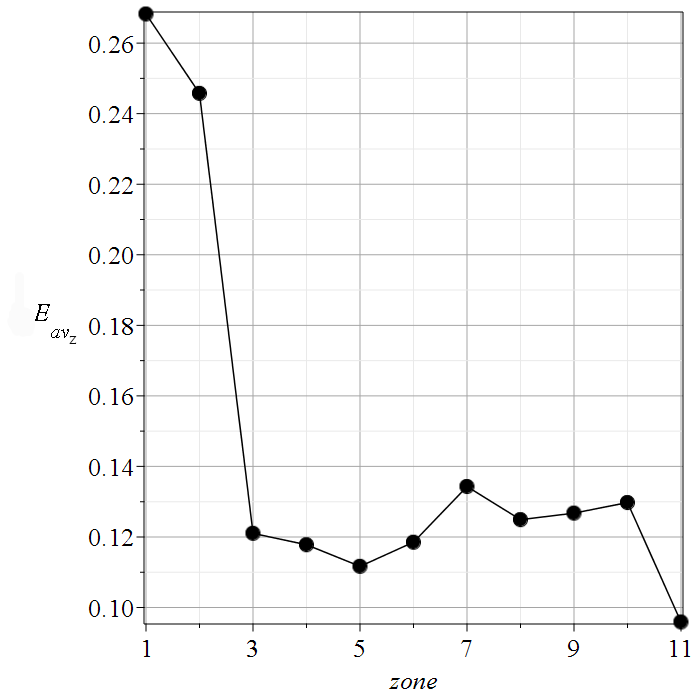}[a]
\includegraphics[scale=0.2]{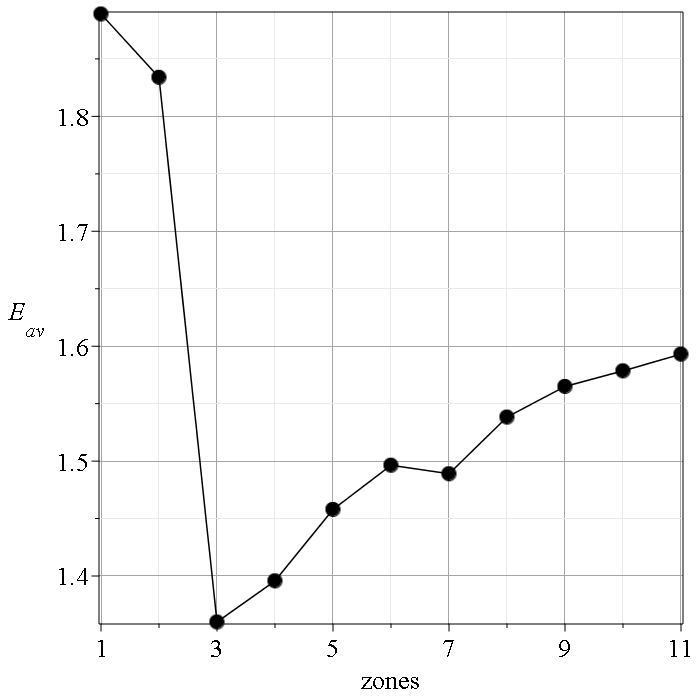}[b]
\caption{(a) The contributions of Bohmian particles in the successive zones ($1-11$) to the value of $E_{av}$  at $t=0$ (b) The separate values of  $E_{av}$ in the various zones.}\label{zones}
\end{figure}

Then we separated the initial Born distribution of Fig.~\ref{triplesin} into various zones of extent $\Delta P=0.03$. Namely, the zones 1,2,...11 contain the particles with P in the intervals $0<P<0.03,$ $0.03<P<0.06,...$ up to $0.3<P<0.33$. We calculated the contribution of the various zones in finding $E_{av}$ and in Fig.~\ref{zones}a we give $E_{{av}_z}$, i.e. the energy of the particles in every zone divided by the total number $N=5000$ of particles. We see that the zones 3,4,..10 contribute roughly equal amounts in $E_{av}$ namely $0.123\pm 0.015$. The last zone (11) contributes a little less (0.096) and the two first zones contribute about double  the average (0.268 in zone 1 and 0.246 in zone 2). In these two zones the contribution of the particles of the second blob is manifest.

If we take the average energy $E_{av}$ in the various zones $1,2,\dots 11$ by dividing their contribution by the corresponding number of particles (and not by $5000$) we see that the average value of the last zone is $1.593$, very close to the global average (Fig.~\ref{zones}b). The previous zones  ($10,9,\dots 3$) have decreasing values of $E_{av}$ except for the first two zones (2 and 1) which have very large values because of the second blob.

On the other hand, the chaotic trajectories are very close to the bottom of the distribution $(P=0)$ and their average values of $E_{av}$ is $1.38$, smaller than the analytical value and close to the minimum value of Fig.~\ref{zones}b. It is notable  that the chaotic trajectories do not give average values close to the analytical value. On the contrary the best approach is provided by the ordered trajectories on the top of the large blob of Fig.~\ref{triplesin}a. Therefore the claim that the distribution of particles tends to the Born distribution because of the chaotic trajectories is not valid in the present case,  first because the number of the chaotic trajectories is small and   second because the chaotic trajectories in the present case do not give the correct average value of the energy. Namely, if we start with an initial distribution of only chaotic trajectories we do not reach the Born distribution in the long run. In fact, even  if we keep the correct proportions of ordered/chaotic trajectories, we cannot reach the Born distribution unless the ordered trajectories have the distribution found within the Born rule. E.g. if we take initially ordered trajectories  only from the top of the large blob we will not ever reach the Born distribution.

Besides the values of the energy it is of interest to find the average values of other basic quantum observables as well, such as the momentum, angular momentum and position.

The average value of the $p_x$ component of the momentum is given analytically by
\begin{equation}\label{pxint}
\langle p_x\rangle=\int_{-\infty}^{\infty}\int_{-\infty}^{\infty}\Psi^{*}\hat{p}_x\Psi dxdy,
\end{equation}
where $\hat{p}_x=-i\hbar\frac{\partial }{\partial x}$ is the corresponding momentum operator and $\Psi^*$ is the complex conjugate of $\Psi$. The analytical calculation of $\langle p_x\rangle$ is given in the Appendix. There we show the various terms of the function $\Psi^{*} \hat{p}_x\Psi$ and indicate those terms that give nonzero contributions in the integral \eqref{pxint}. In the present case we find after some algebra that
\begin{equation}\label{pxms}
\langle p_x\rangle=-\frac{\sqrt{2\omega_x}ab\sin(\omega_xt)}{a^2+b^2+c^2}.
\end{equation}
and similarly
\begin{equation}\label{pyms}
\langle p_y\rangle=-\frac{\sqrt{2\omega_y}bc\sin(\omega_yt)}{a^2+b^2+c^2}.
\end{equation}

\begin{figure}[H]
\centering
\includegraphics[scale=0.23]{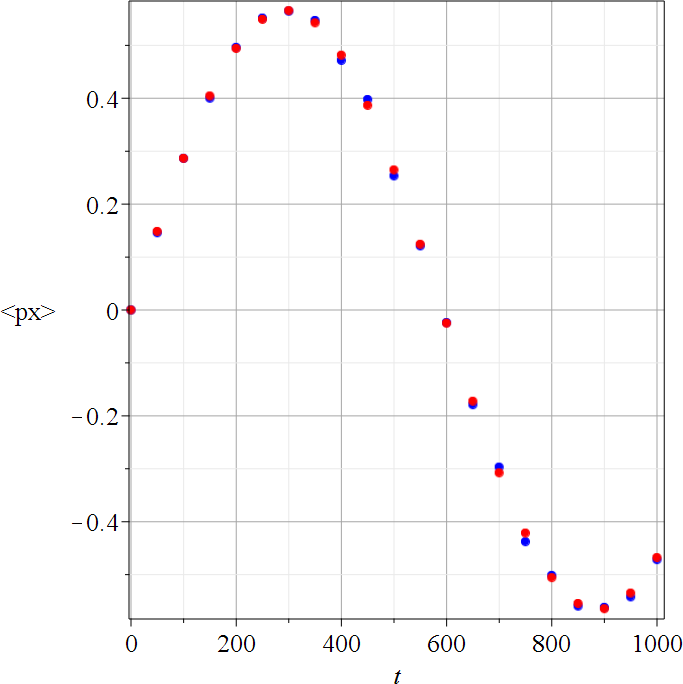}[a]
\includegraphics[scale=0.23]{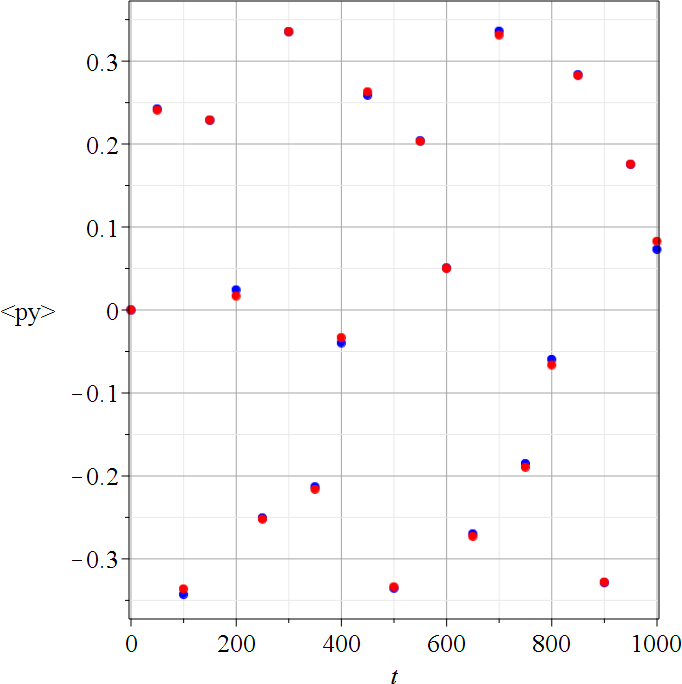}[b]
\caption{The average values of $\langle p_x\rangle$ (a) and $\langle p_y\rangle$ (b) as  functions of time for $t\in[0,1000]$ for every $\Delta t=50$. The red dots correspond to the analytical values produced {by Eqs.~\ref{ormes} }and the blue dots correspond to the values produced by  5000 Bohmian particles distributed according to BR. {We observe that, in both cases, the differences between the analytical and the numerical values are very small.}}\label{pxpy}
\end{figure}

Therefore $\langle p_x\rangle$ and $\langle p_y\rangle$ are time periodic with periods $2\pi$ and $2\pi\sqrt{2}$ respectively.
In Fig.~\ref{pxpy}a we give the values of $\langle p_x\rangle$, which change with a period $T=2\pi$, up to $t=1000$. We have marked with red dots the values of $\langle p_x\rangle$ at every multiple of $t=50$. In every interval $\Delta t=50$ take place about 8 oscillations. If we calculate now the average values of $\langle p_x\rangle_{av}$  of 5000 particles at these times we find almost exactly the same values (blue dots) with a mean deviation in $p_x$ equal to $\sim 0.005$, i.e. of order 1\%. Similar values are   found in Fig.~\ref{pxpy}b  for $\langle p_y\rangle$. The mean deviation in $p_y$ is  $\sim 0.003$.

The average angular momentum is given analytically by
\begin{eqnarray}
\nonumber \langle L\rangle &=
\int_{-\infty}^{\infty}\int_{-\infty}^{\infty}(\Psi^{*}x\hat{p_y}-y\hat{p}_x)\Psi dxdy\\&={
\frac {1}{\sqrt {\omega_x\omega_{y}}}}\frac {ac \left( \omega_{x}-\omega_{y} \right) \sin \left( \left( 
\omega_{x}+\omega_{y} \right) t \right) }{{a}^{2}+{b}^{2}+{c}^{2}}\label{Lms}
\end{eqnarray}
and  similarly the average values $\langle x\rangle, \langle y\rangle$ are found equal to (see the Appendix for details):
\begin{align}\label{xms}
\langle x \rangle =\sqrt{\frac{2}{\omega_x}}\frac{ab\cos(\omega_x t)}{a^2+b^2+c^2}\\
\langle y \rangle =\sqrt{\frac{2}{\omega_y}}\frac{ab\cos(\omega_y t)}{a^2+b^2+c^2}\label{yms}.
\end{align}

We note that 
\begin{equation}
\langle p_x\rangle=\frac{d\langle x\rangle}{dt}, \, \langle p_y\rangle=\frac{d\langle y\rangle}{dt}.\label{ormes}
\end{equation} 

In the present case for $t=0$ we have $\langle x
\rangle= 0.566$ and $\langle y\rangle=0.336$ while $\langle p_x(0)\rangle=\langle p_y(0)\rangle=0$. The mean deviations between the values from SQM and the numerical values from our Bohmian simulation of $\langle L\rangle$, $\langle x\rangle, \langle y\rangle$ for $t\in [0,1000]$  are very small (see Tab.~\ref{tab1}).

For any arbitrary distribution of particles $P_0\neq|\Psi_0|^2$ the mean deviations of the numerical average values from the analytical ones are in general large (see Table~\ref{tab1}).

\begin{figure}[H]
\centering
\includegraphics[scale=0.25]{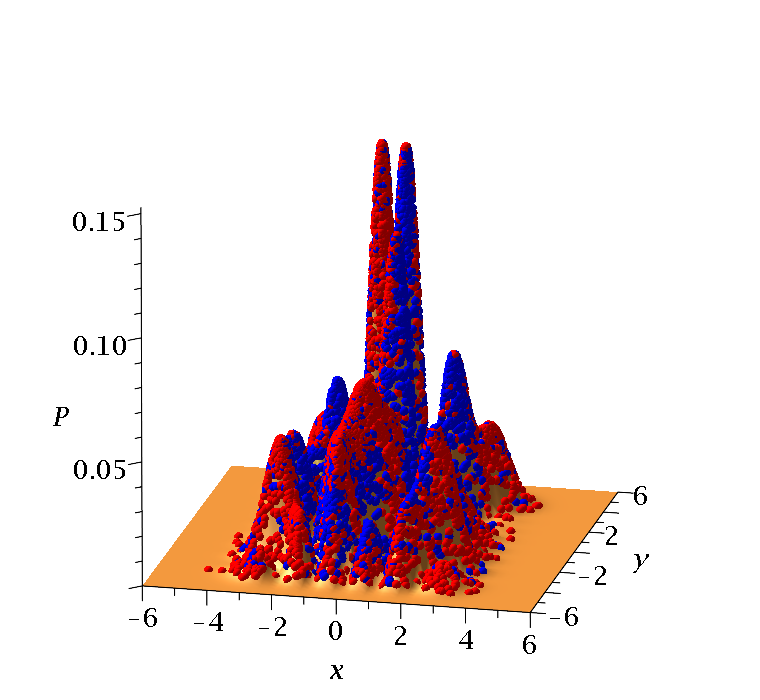}[a]
\includegraphics[scale=0.18]{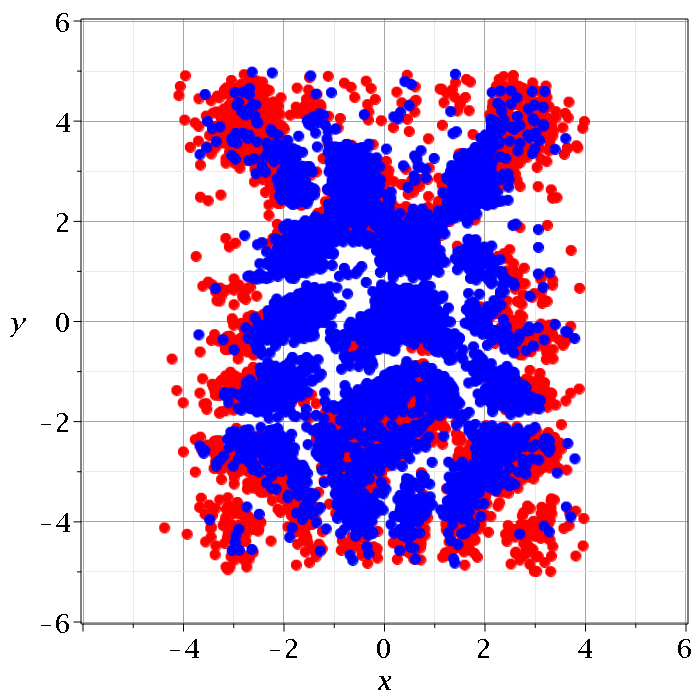}[b]\\
\includegraphics[scale=0.20]{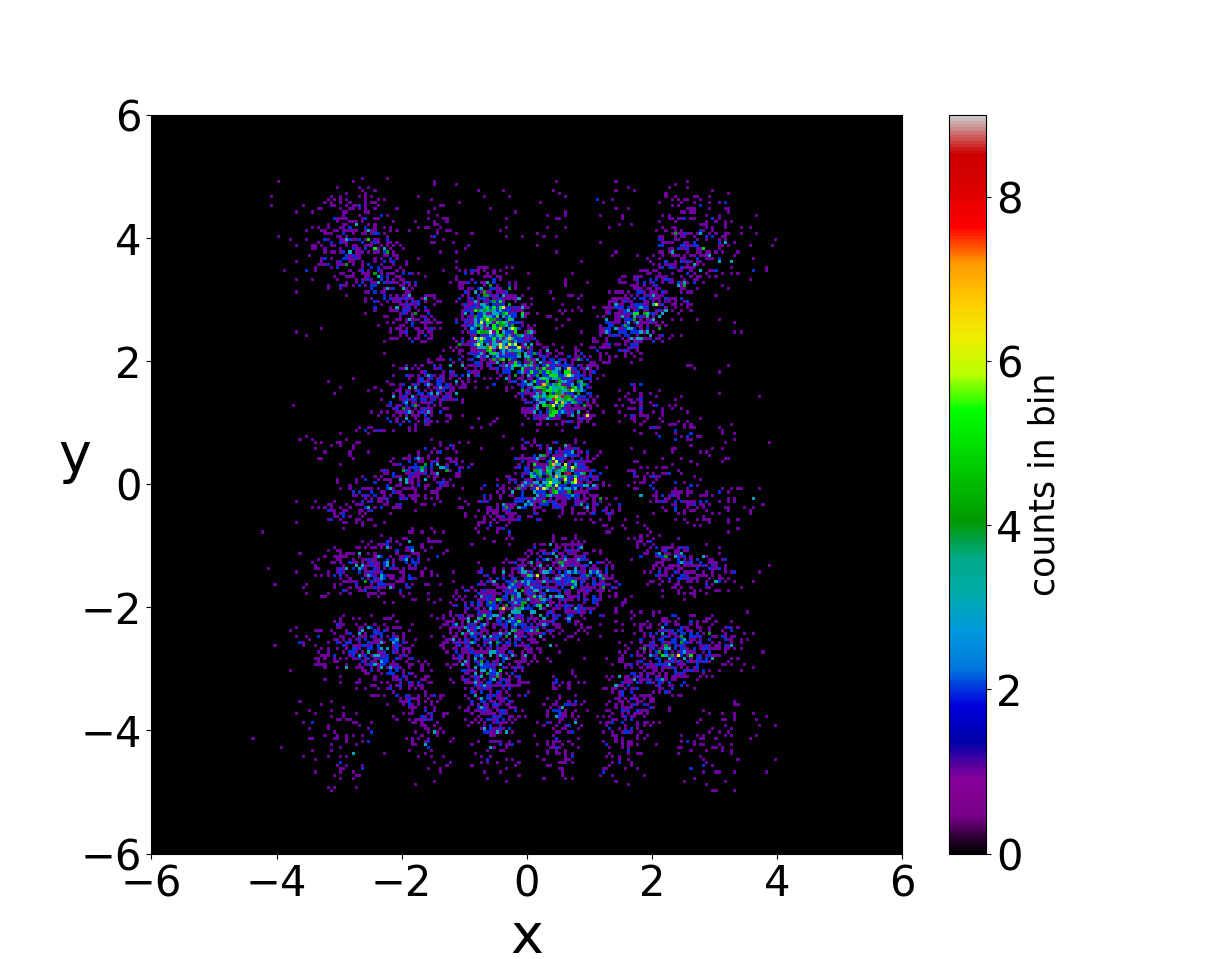}[c]
\includegraphics[scale=0.185]{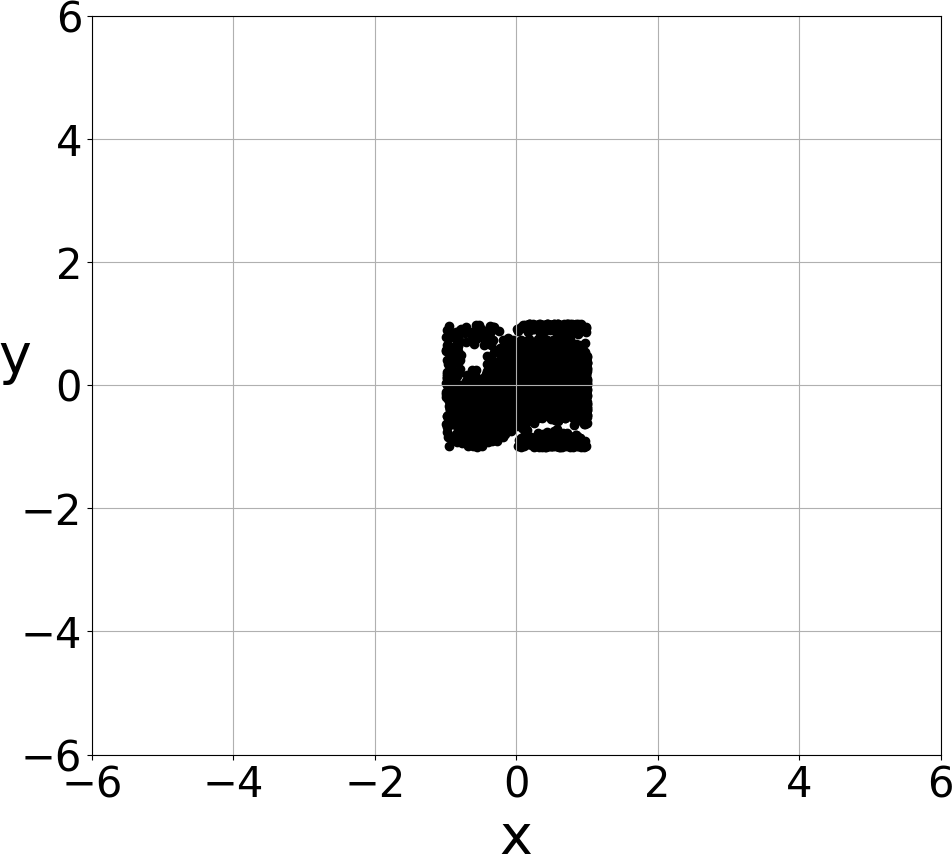}[d]
\caption{a,b,c: 10000 initial conditions of ordered (red) and chaotic (blue) trajectories distributed according to BR in the case of the wavefunction $\Psi=M(a\Psi_{0,2}+b\Psi_{3,4}+c\Psi_{5,7})$: a) on the surface of $P=|\Psi|^2$  and b) on the $x-y$ plane. In c) we see their colorplot at $t=0$. d) 10000 non-Born distributed initial conditions.}\label{10000}
\end{figure}

\section{Wavefunctions with many nodal points}

\subsection{Case 1}
A wavefunction with many nodal points is 

\begin{equation}
\Psi=M(a\Psi_{0,2}+b\Psi_{3,4}+c\Psi_{5,7}),
\end{equation}
where the functions $\Psi_{m,n}$ are given by Eq.~\ref{mn}. The constants have the same values as in Section 2.

 In this case the maximum $m_i$ is $5$ and the maximum $n_i$ is $7$, therefore the number of the nodal points is of order 35. An initial realization of Born's rule made out of $10000$ Bohmian particles along with the surface $P_0=|\Psi_0|^2$ is shown in Figs.~\ref{10000}a,b. We observe that $P$ has many blobs that collide as they evolve in time \cite{tzemos2021role}. The proportions of ordered (red) and chaotic (blue) trajectories are  in this case about 50\%, i.e. almost equal. The distribution of the particles at $t=0$ is shown in the initial colorplot (Fig.~\ref{10000}c).

\begin{figure}
\centering
\includegraphics[scale=0.2]{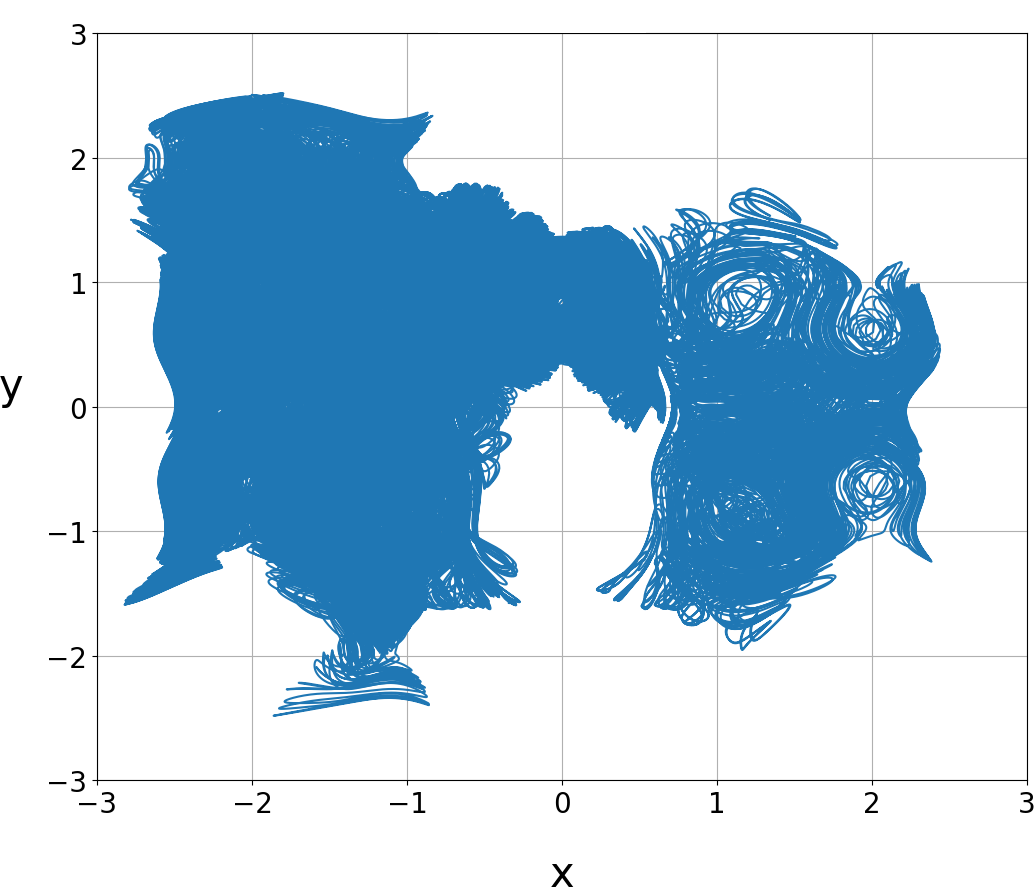}[a]
\includegraphics[scale=0.2]{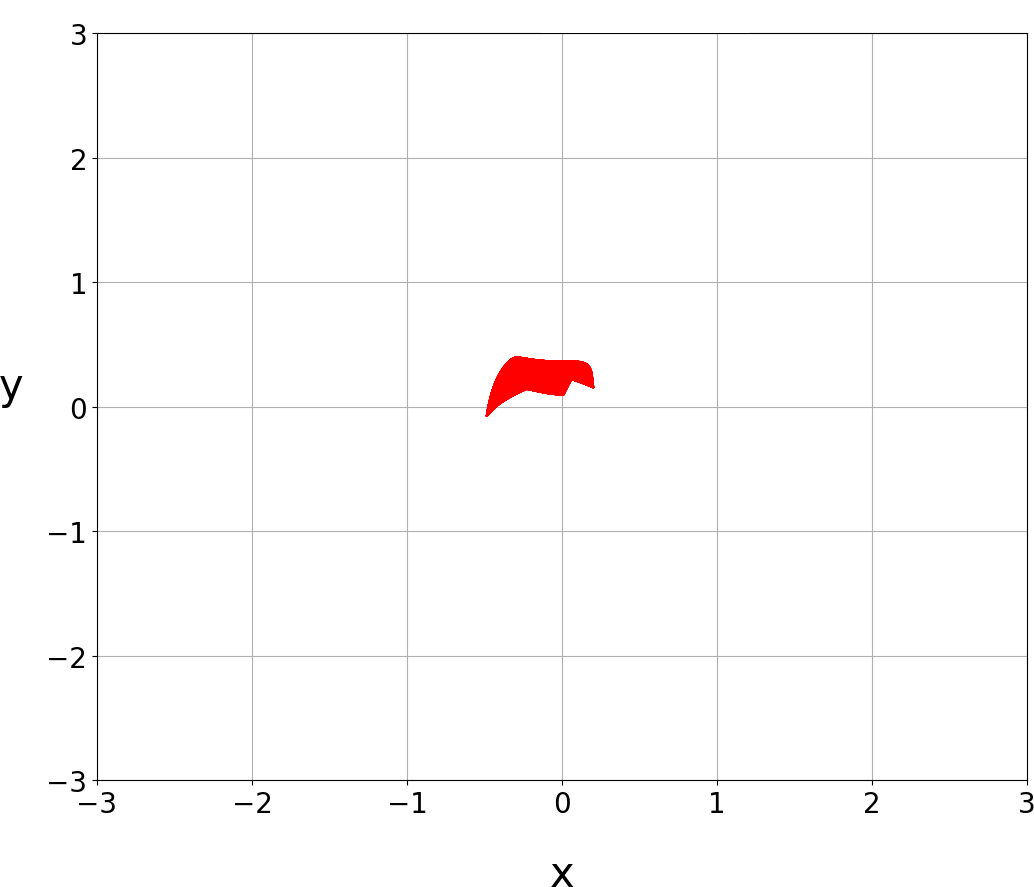}[b]
\caption{a,b) A chaotic $(x(0)=-0.6, y(0)=-0.01)$ (a), and an ordered ($x(0)=0.01, y(0)=0.1$)  trajectory (b), of the wavefunction $\Psi=M(a\Psi_{0,2}+b\Psi_{3,4}+c\Psi_{5,7})$. }\label{orc}
\end{figure}

\begin{figure}
\centering
\includegraphics[scale=0.2]{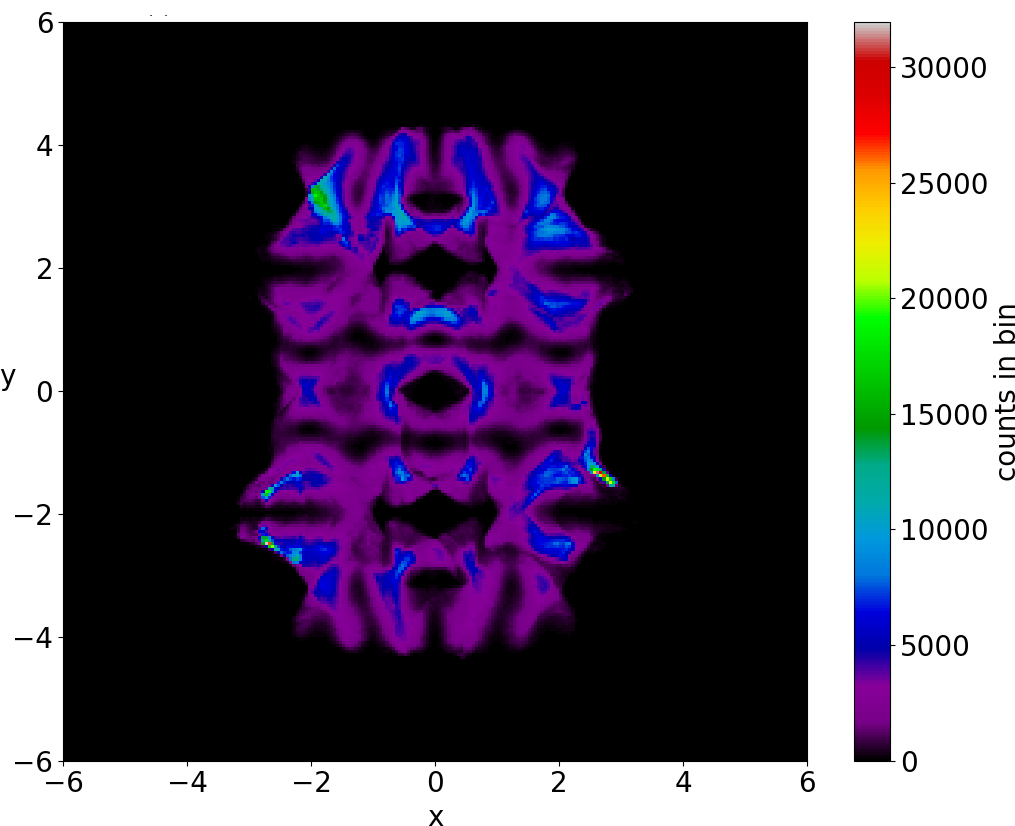}[a]
\includegraphics[scale=0.2]{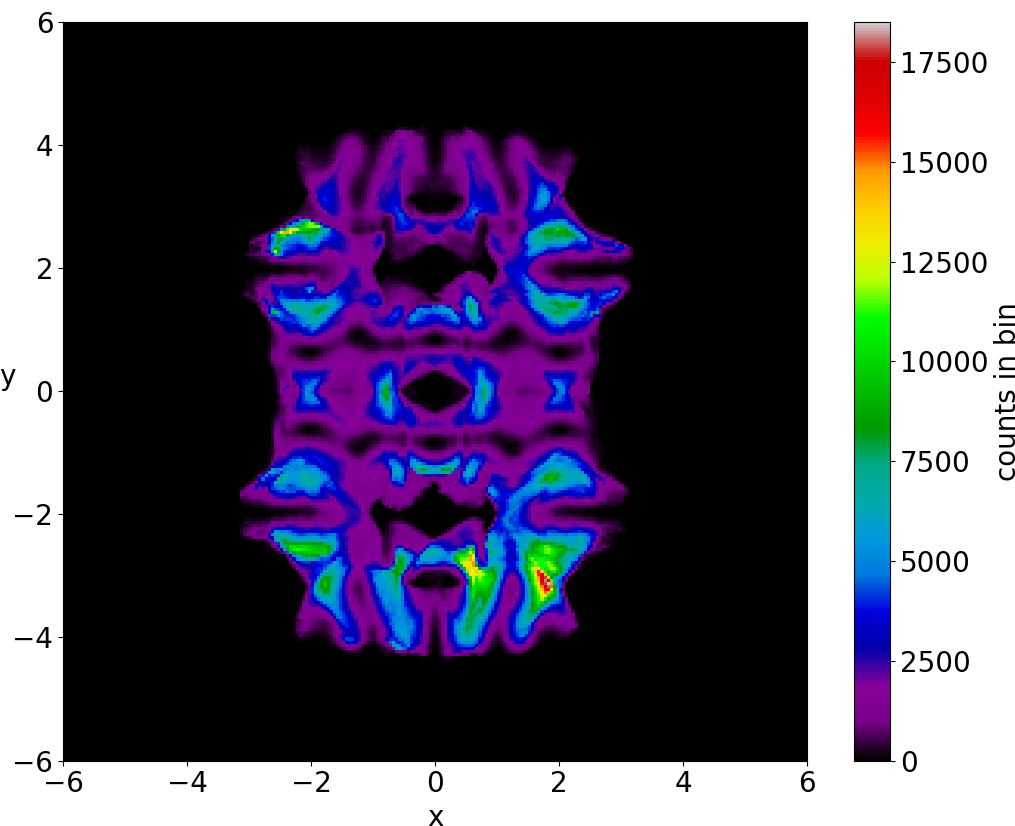}[b]
\caption{The colorplots of two chaotic trajectories of the wavefunction $\Psi=M(a\Psi_{0,2}+b\Psi_{3,4}+c\Psi_{5,7})$ up to $t=2\times 10^6$:  a) $x(0)=0.6693, y(0)=-2.4559$ b) $x(0)=-0.0631, y(0)=-1.4241$. 
}\label{frob1}
\end{figure}

We observe that the ordered trajectories are mainly close to the tops of the two main blobs. This supports our previous results on qubit states \cite{tzemos2021role,tzemos2020ergodicity} made of coherent states of the quantum harmonic oscillator, where we found that the concentration of the ordered trajectories in the Born distribution is at the top of the larger of the two blobs of $|\Psi|^2$.

The forms of the the chaotic trajectories are very irregular (Fig.~\ref{orc}a), while the  ordered trajectories are like distorted Lissajous figures (Fig.~\ref{orc}b). Moreover the chaotic trajectories are ergodic as we can see in Figs.~\ref{frob1}a,b  where we show two different chaotic trajectories having very similar colorplots (cumulative distributions of points taken at every $\Delta t=0.05$ up to $t=2\times 10^6$).

\begin{figure}[H]
\centering
\includegraphics[scale=0.2]{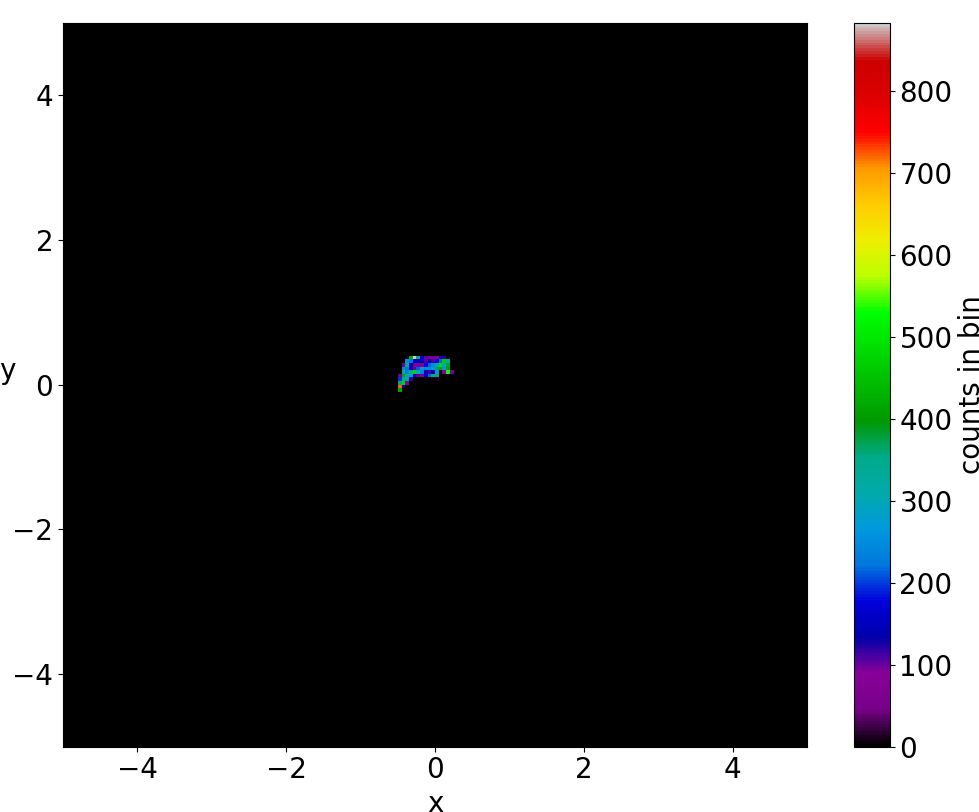}
\caption{The colorplot of an ordered trajectory ($x(0)=0.01, y(0)=0.1$) up to $t=1000$. We see that it covers a very small area of the configuration space.}\label{order}
\end{figure}

\begin{figure}[H]
\centering
\includegraphics[scale=0.18]{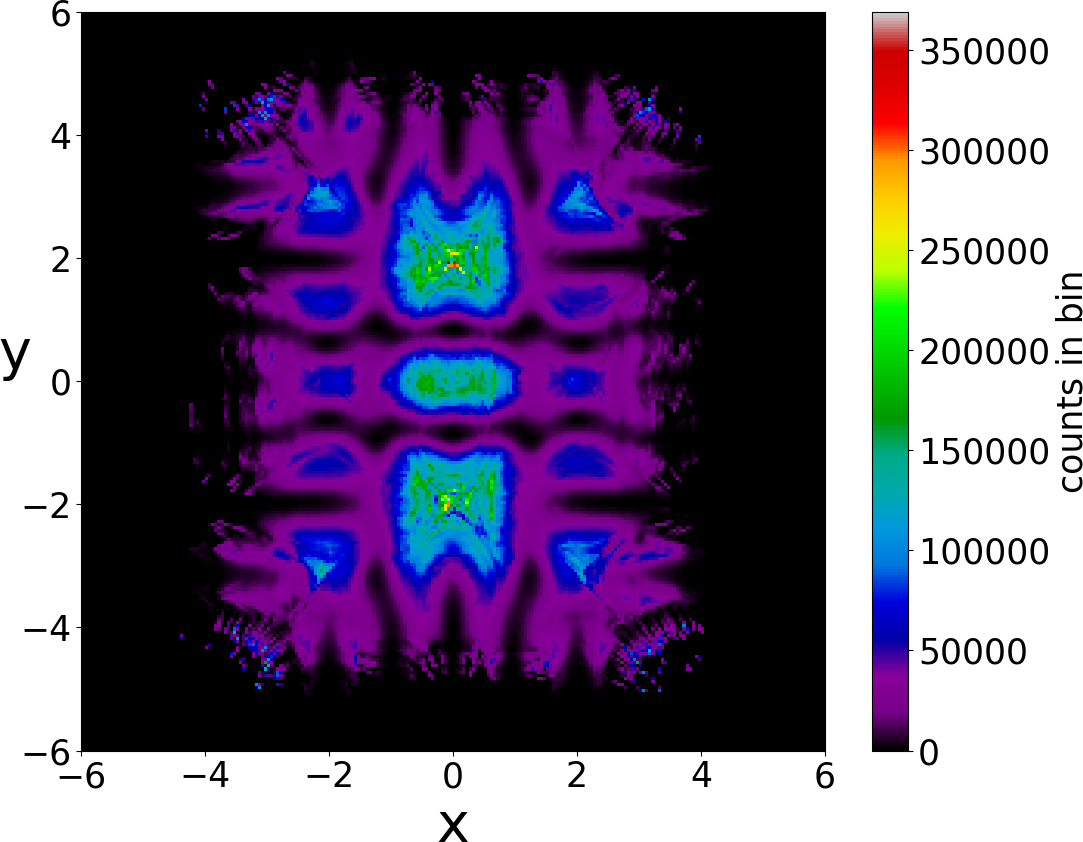}[a]\\
\includegraphics[scale=0.18]{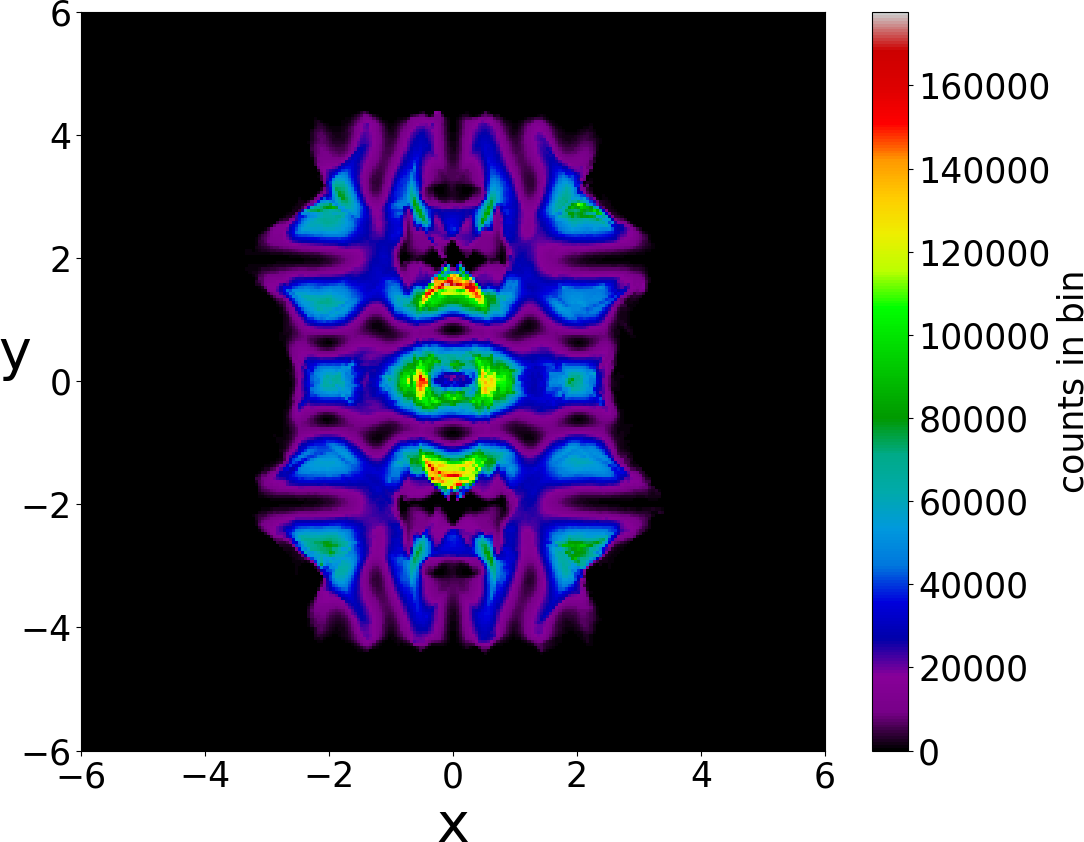}[b]
\includegraphics[scale=0.18]{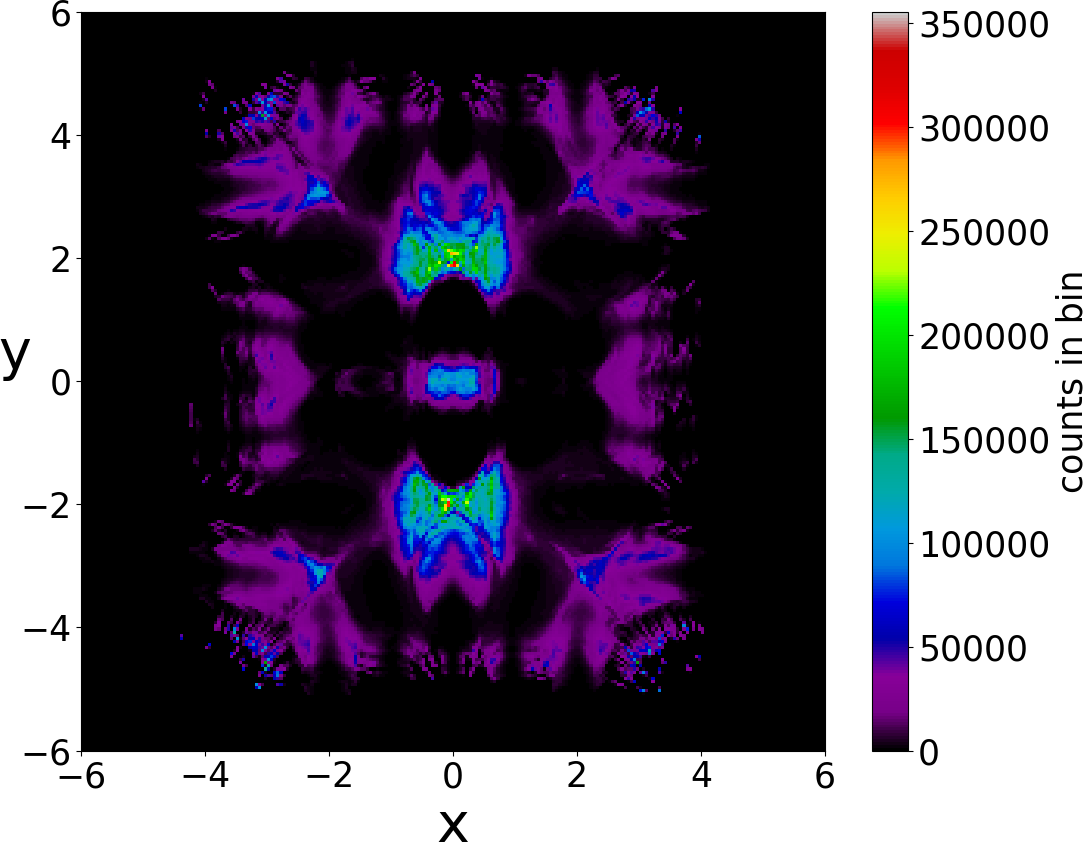}[c]
\caption{a) The colorplot of a realization of BR distribution consisting  of $10000$ particles integrated up to $t=5000$ in the case of the wavefunction $\Psi=M(a\Psi_{0,2}+b\Psi_{3,4}+c\Psi_{5,7})$. The colorplots of the  $5053$ chaotic and  $4947$ ordered  trajectories are given separately in (b) and (c).}\label{triplemult1}
\end{figure}

On the other hand the colorplots of ordered trajectories cover small parts of the space $(x,y)$ (Fig.~\ref{order}). Therefore their colorplots are very different from those of the chaotic trajectories.

In Fig.~\ref{triplemult1}a we show the  colorplot of all the particles up to a time $t=5000$, while the colorplots of the chaotic and of the ordered trajectories up  to this time are shown in Figs.~\ref{triplemult1}b,c. We see that the colorplots  Figs.~\ref{triplemult1}a,b,c are quite different from each other.  Therefore, if we take only chaotic (or only ordered) trajectories we will never reach the Born distribution.

As regards the average values of the observed quantities in this case we have the following: From SQM the average energy is 
\begin{equation}
\langle E\rangle=M^2(|a|^2E_{0,2}+|b|^2E_{3,4}+|c|^2E_{5,7}),
\end{equation}
where $E_{m,n}$ are given by Eq.~\ref{ener}. Using the values $a=b=1,c=\sqrt{2}/2$ and $M^2=(a^2+b^2+c^2)^{-1}$ we find that $\langle E_{av}\rangle=5.7406$. Then we calculate the average value $E_{av}$ of $N=10000$ Bohmian particles  we find that, at $t=0$, $E_{av}=5.7438.$ Therefore the initial  deviation  is of order $0.003$, i.e.  $0.06\%$. 

The average values of the momentum $\langle p_x\rangle, \langle p_y\rangle$, of the angular momentum $\langle L\rangle$ and of the position $\langle x\rangle$ $\langle y\rangle$ are calculated analytically in the Appendix. In the present case they are all zero, i.e., $\langle p_x\rangle=\langle p_y\rangle=\langle L \rangle=\langle x\rangle=\langle y\rangle=0$. In fact, in the Appendix we show that they are zero whenever $m$ and $n$ differ by more than 1 (i.e. whenever they are not adjacent integers).  The corresponding mean absolute deviations from $\langle p_x\rangle, \langle p_y\rangle$ for  $10000$ Born distributed Bohmian particles and $t\in[0,1000]$ are  $0.009$ and  $0.007$  respectively. Moreover, the deviations from the average angular momentum $\langle L\rangle$ and the average positions $\langle x\rangle, \langle y\rangle$ are $0.0234$, $0.0122$ and $0.0158$  (see Table~\ref{tab1}) and they tend to zero as time increases considerably.

On the other hand, in a case  with $10000$ particles distributed randomly in the  square $(-1,1)\times(-1,1)$ , i.e. away from the BR distribution (Fig.~\ref{10000}d), we found  much larger deviations for the same  time, namely  about  $10$ times larger than the values of the Born distribution (see Table~\ref{tab1}),  which do not decrease as $t$ increases.

\subsection{Case 2}

If we take 
\begin{align}
\Psi=M(a\Psi_{10,3}+b\Psi_{4,5}+c\Psi_{7,8})
\end{align} we have $m_i^{max}=10, n_i^{max}=8$, therefore the number of the nodal points is of order $80$. In this case the proportion of chaotic trajectories is about $96.2\%$ and the ordered trajectories are only $3.8\%$.

\begin{figure}[H]
\centering
\includegraphics[scale=0.3]{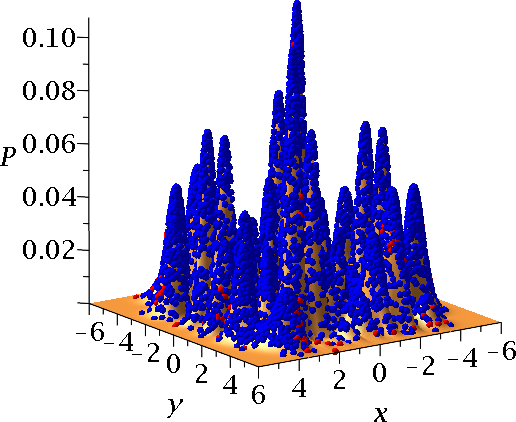}[a]
\includegraphics[scale=0.2]{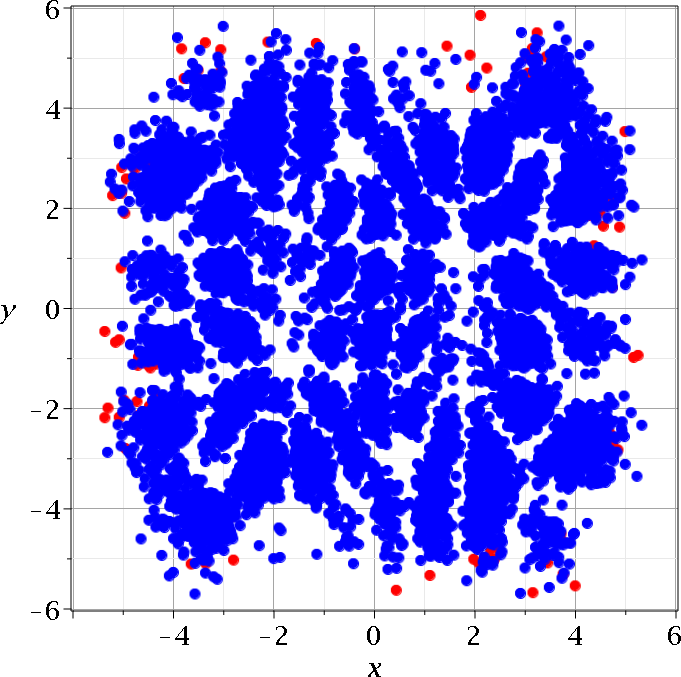}[b]
\includegraphics[scale=0.2]{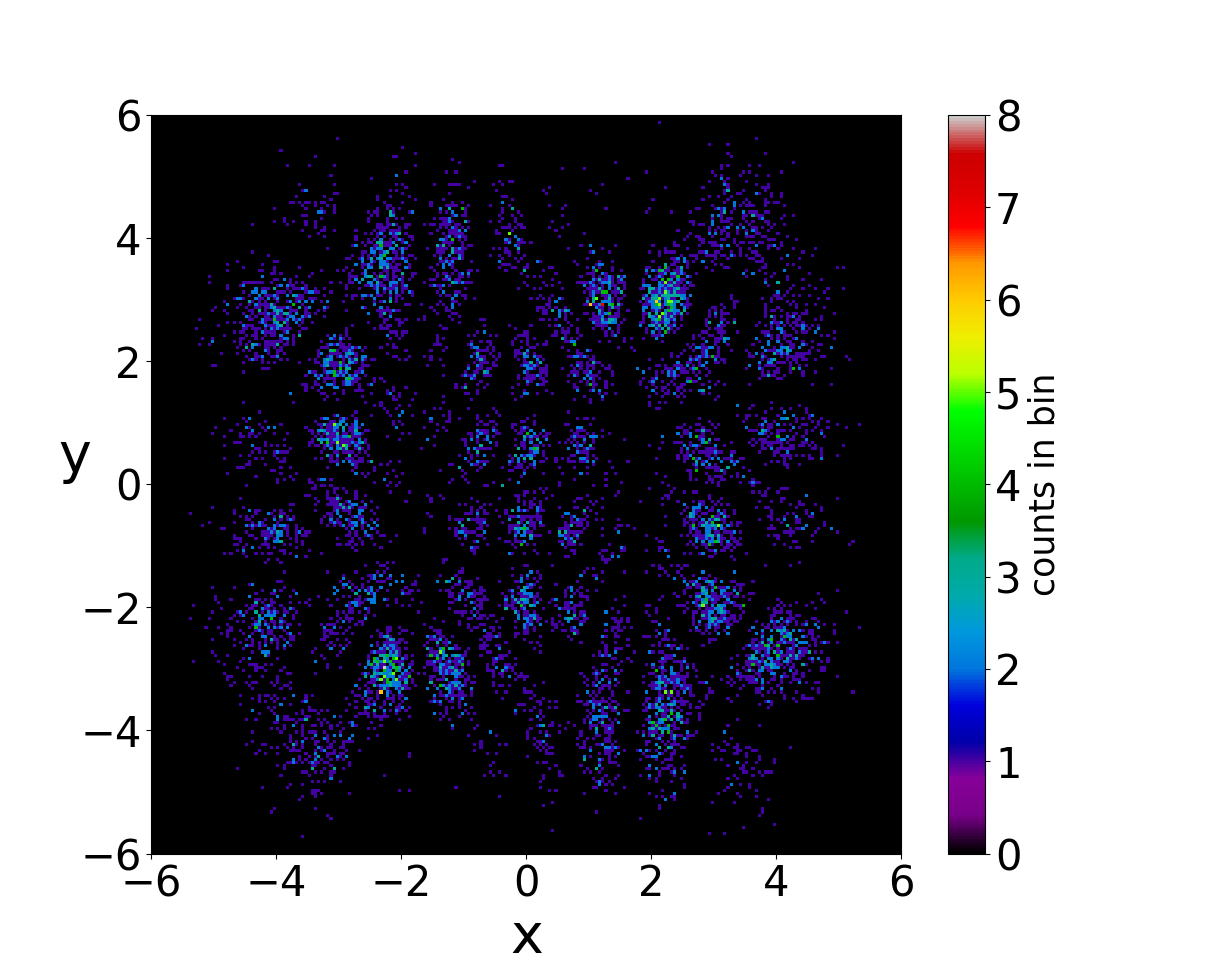}[c]
\caption{10000 initial conditions of ordered (red) and chaotic (blue) trajectories distributed according to BR in the case of the wavefunction $\Psi=M(a\Psi_{10,3}+b\Psi_{4,5}+c\Psi_{7,8})$: a) on the surface of $P_0=|\Psi_0|^2$  and b) on the $x-y$ plane ($t=0$). c) Their colorplot at $t=0$.}\label{1034578}
\end{figure}

\begin{figure}[H]
\centering
\includegraphics[scale=0.2]{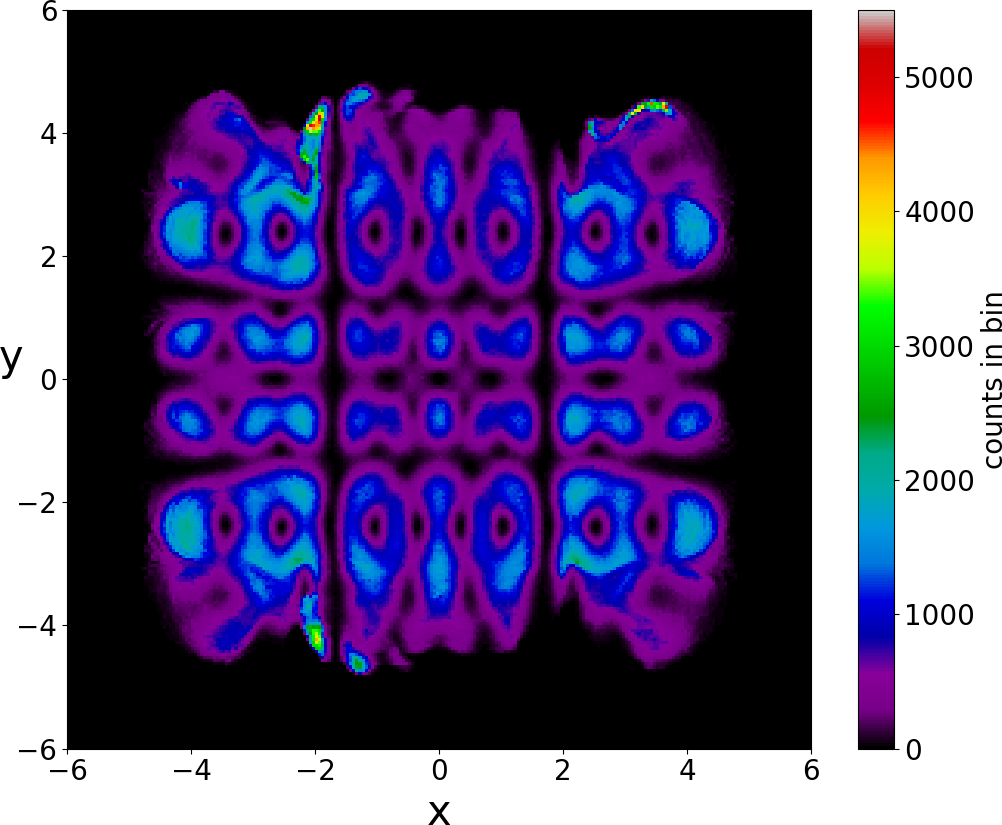}[a]
\includegraphics[scale=0.2]{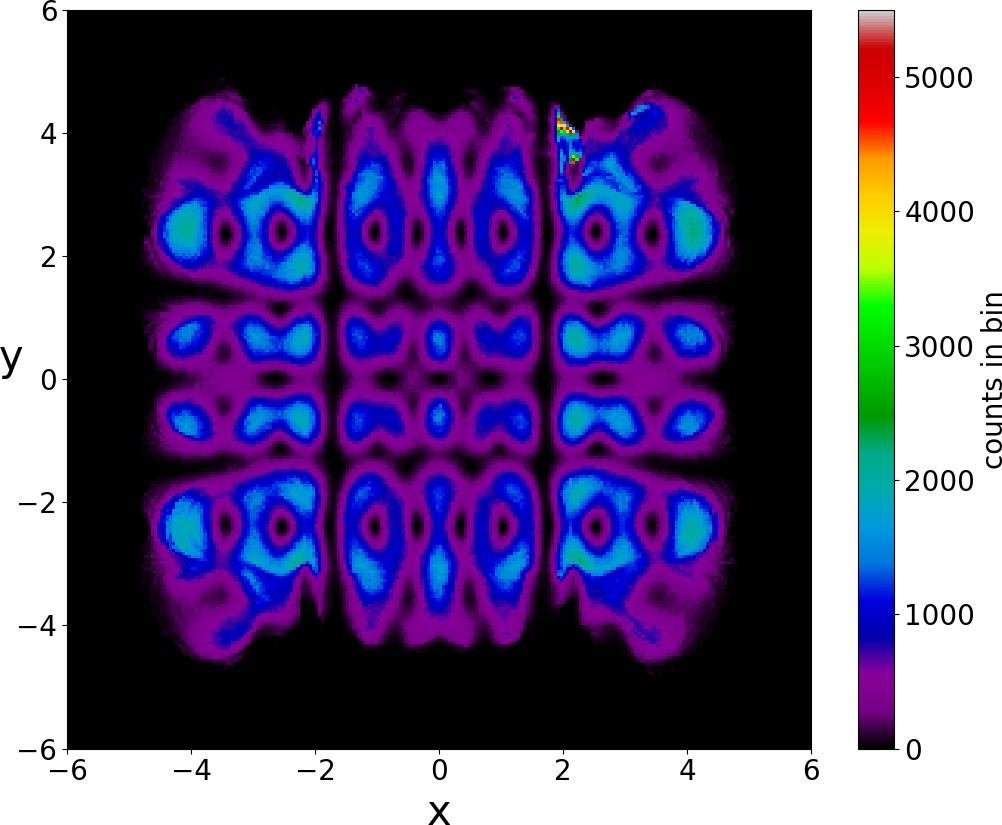}[b]
\caption{The colorplot of two chaotic trajectories up to $t=10^6$ in the case $\Psi=M(a\Psi_{10,3}+b\Psi_{4,5}+c\Psi_{7,8})$: a) $x(0)=-1 ,y(0)=0.7$ and b) $x(0)=-0.1, y(0)=-1.5$. They look very similar.}\label{1034578_c12}
\end{figure}

The initial distributions of the points according to BR is given in Fig.~\ref{1034578}. In this case the tops of the main blobs contain mainly chaotic trajectories, which in fact are ergodic. This is shown in Fig.~\ref{1034578_c12} where we compare the colorplots of two chaotic trajectories up to $t=10^6$.

\begin{figure}[H]
\centering
\includegraphics[scale=0.25]{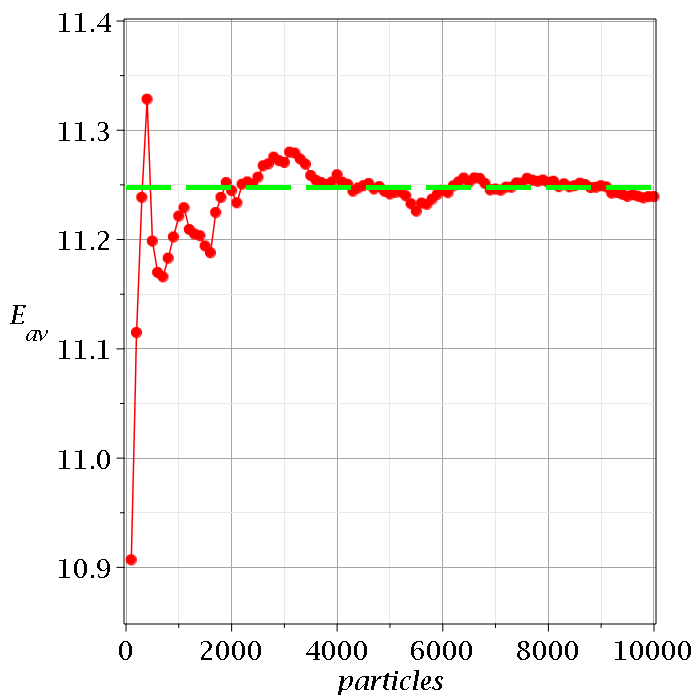}
\caption{The average value of energy as a function of the number of trajectories in the realization of Born's rule at $t=0$.}\label{1034578en}
\end{figure}

In particular, the theoretical average energy is $\langle E\rangle= M^2(a^2E_{10,3}+b^2E_{4,5}+|c|^2E_{7,8})$ and this is equal to $\langle E\rangle=11.248$. In Fig.~\ref{1034578en} we show the numerical  average at $t=0$ as a function of the number of Bohmian particles in the Born distribution. We see that the numerical value converges to $E_{av}=11.239$  when the number $N$ is larger than $2000$. The errors are about $ 0.009$, i.e. of order $0.08\%$.

The analytical values of the  average momenta $\langle p_x\rangle$ and $\langle p_y\rangle$, angular momentum $\langle L\rangle$  and position $\langle x\rangle, \langle y\rangle$ are zero (since the values of $m$ and $n$ are not different by 1, see the Appendix).

The mean absolute deviations in time are  also found to be very close to zero. In particular the deviations $|\delta \langle p_{x}\rangle|_{av},|\delta \langle p_{y}\rangle|_{av}$ for 10000 particles following Born's distribution and for $t\in[0,1000]$ are 0.0082 and 0.0081. In this case the variations of a random non-Born distribution in the square $(-1,1)\times (-1,1)$ (taken the same as in the case 3.1) are  $0.011$ and $0.013$, i.e. of the same order as for an initially Born distribution  (see Table~\ref{tab1}). This is due to the fact that any non-Born distribution tends to reach BR distribution in the long run when it is dominated by chaotic-ergodic trajectories.

\section{Conclusions}

In this paper we have considered mainly distributions of particles that satisfy Born's rule $P=|\Psi|^2$. In general Born's distribution contains both chaotic and ordered trajectories. The chaotic trajectories are ergodic.

In the case of one nodal point the ordered trajectories are dominant. In the first case of many nodal points the ordered trajectories are almost equal to  the chaotic trajectories, while in the second case the chaotic trajectories are dominant. In this case the Born distribution is essentially satisfied in the long run, even if the initial distribution is not  Born-distributed.

However the existence of many nodal points does not necessarily imply that chaos is dominant.
In fact, there are special cases of multinodal wavefunctions which are  dominated by ordered trajectories. Nevertheless these cases are rather exceptional.

We calculated analytically the average values of basic quantum  quantities, namely  the energy $\langle E\rangle$, the momentum $\langle p_x\rangle, \langle p_y\rangle$, the angular momentum $\langle L\rangle$ and the position $\langle x\rangle,\langle y\rangle$. Their values are found analytically in the Appendix for systems of three components $\Psi=M(a\Psi_{m_1,n_1}+b\Psi_{m_2,n_2}+c\Psi_{m_3,n_3})$.  We found the necessary conditions in order to have nonzero values of $\langle p_x\rangle, \langle p_y\rangle, \langle L\rangle, \langle x\rangle, \langle y\rangle$. E.g. in calculating $\langle p_x\rangle$ we must have at least two $m_i$ indices to be different by 1 and the corresponding $n_i's$ equal.
Similarly in calculating $\langle p_y\rangle$ we must have at least two $n_i$ indices to be different by 1 and the corresponding $m_i's$ equal. In the present case of one nodal point these conditions are satisfied and $\langle p_x\rangle , \langle p_y\rangle$, are periodic in time.  In general these conditions are not satisfied and the values of $\langle p_x\rangle, \langle p_y\rangle$ are zero. Similar conditions exist for the average values of the other quantum observables.

We compared the analytical values with the average values of the observed quantities by calculating a large number of trajectories with a Born initial distribution using the Bohmian equations of motion. {As expected,} we found a good agreement with the analytical values of SQM when the number of trajectories is sufficiently large. 
\begin{table}
  \begin{tabular}{|c|c|c|c|c|c|c|c|}
    \hline
    Nodal points& \multicolumn{2}{c|}{One}& \multicolumn{4}{c|}{Many}\\ 
   \hline
   Distribution& Born(3/97)& Non-Born& Born(50-50)& Non-Born& Born(96/4)&Non-Born\\
   \hline
    $|\delta\langle E\rangle|_{av}$&0.0047& 0.1942&0.0348 & 1.9518& 0.0171& 0.1338\\
   \hline
   $|\delta\langle p_x\rangle|_{av}$& 0.0053& 0.2026& 0.0094& 0.2920& 0.0082& 0.0107\\
   \hline
    $|\delta\langle p_y\rangle|_{av}$& 0.0032& 0.1525&0.0072&0.2265& 0.0081& 0.0130\\
   \hline
      $|\delta\langle L\rangle|_{av}$& 0.0082& 0.3019&0.0158& 0.1501& 0.0416& 0.1113\\
         \hline
    $|\delta\langle x\rangle|_{av}$& 0.0073& 1.0237& 0.0234& 0.2177& 0.0593&0.0896\\
   \hline
   $|\delta\langle y\rangle|_{av}$& 0.0105& 0.2122&0.0122& 0.1421&0.0309&0.0366\\
   \hline

  \end{tabular}
  \caption{Table of mean deviations of quantum observables between the predictions of SQM and our numerical simulations of examples of distributions of Bohmian particles for $t\in[0,1000]$ for every $\Delta t=50$. In parentheses we show the percentages of the chaotic and ordered trajectories in our realizations of Born distribution. We observe in the last column  of the table (non-Born distribution in the case 3.2) that the mean deviations of the energy and  of the angular momentum are still relatively large. They decrease for larger integration times, because in this case the  distribution of particles tends to the BR distribution. }\label{tab1}
\end{table}

However, in the cases with appreciable numbers of ordered trajectories the average numerical results do not agree with the analytical values if the initial distribution does not follow the Born rule. This is shown in Table~\ref{tab1} where we compare the errors of the average quantities with respect to the analytical values in the Born and the non-Born cases. The errors in the Born case are small and become even smaller when the integration time increases. On the other hand,  the errors of the non-Born case are in general much larger and do not decrease in general when the time increases,  except in the second case of many nodal points when the chaotic-ergodic trajectories are dominant. In this particular case any initial non-Born distribution tends practically to Born's distribution after a large time. 
\begin{figure}[H]
\centering
\includegraphics[scale=0.25]{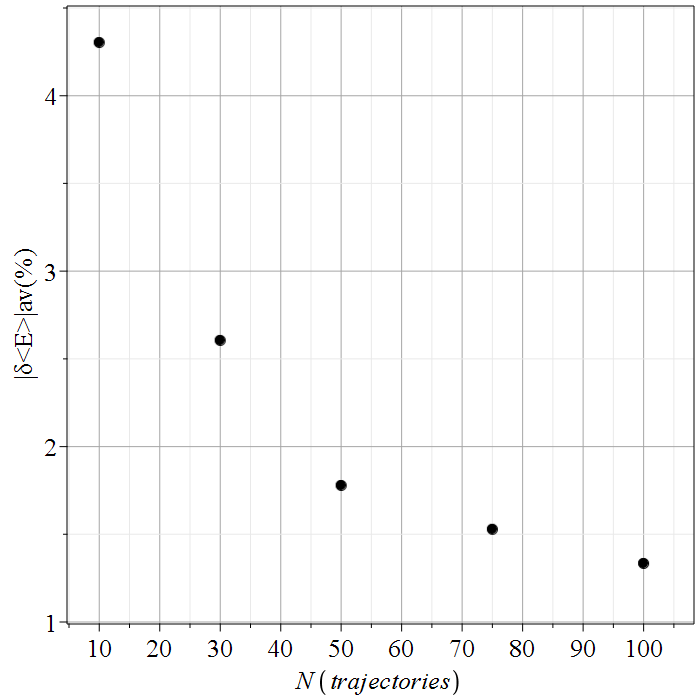}
\caption{The mean absolute deviation (percent) between ensembles of chaotic trajectories in the case of full chaos, for $t\in [0,10^5]$ with a time step $\Delta t=50$ {as a function of the number of the trajectories in the ensemble.} Even with a few trajectories we find an average value of energy rather close to the analytical with deviations of only a few percent.}\label{ligc1034578}
\end{figure}

In fact, in a system where chaos is dominant one needs only a few trajectories in order to calculate with a satisfying accuracy the average values of quantum observables, but at the cost of a large integration time. This can be seen in Fig.~\ref{ligc1034578} where we plot the mean absolute deviations between the theoretical and numerical average energy in the multinodal wavefunction $\Psi=M(a\Psi_{10,3}+b\Psi_{4,5}+c\Psi_{7,8})$,
which is dominated by chaotic trajectories,  as a function of the number of trajectories in our sample. The trajectories are integrated up to $t=10^5$ and are sampled at every $\Delta t=50$. We see that even with a few chaotic trajectories we find quite accurate results. E.g. with only 50 trajectories we find an average error less than $2\%$. These results  become  more accurate if we extend the  integration times.

On the other hand, the ordered trajectories are mainly concentrated near the maxima of the Born distribution. They cover only a small part of the configuration space and their long time shape is not common for all of them. Therefore they cannot be simulated  by the few ordered trajectories that follow Born's rule or  by many ordered trajectories that are not Born-distributed.

Thus far, our series of papers on Bohmian quantum chaos has primarily focused on the generation of chaos in Bohmian trajectories and the impact of quantum entanglement \cite{horodecki2009quantum} on the  chaos-order interplay in Bohmian trajectories. Entanglement, quantified by the von Neumann entropy \cite{tzemos2019bohmian} (the quantum analogue of Shannon's entropy) was found to play a key role  in  the dynamical approximation of Born’s rule by  arbitrary initial particle distributions (for the relation between chaos and Shannon's entropy see \cite{PhysRevE.99.062209}).

In systems of Bohmian qubits made out of coherent states of the harmonic oscillator we showed that chaotic trajectories increase with the degree of entanglement. In the case of maximal entanglement all the trajectories are chaotic (and ergodic) while in the zero entanglement case all trajectories are ordered.  

{In the present work we made a step further by trying to understand the contribution of chaotic and ordered Bohmian trajectories inside a Born distributed ensemble of Bohmian particles in the calculation of the values of the observables, something very important for the field of Bohmian chaos.}

{We note that, up to now, all the works on Bohmian chaos have been focused on quantum systems that evolve according to the Schrö\-dinger equation, i.e. closed quantum systems. This implies that the effects due to the environment, such as decoherence and dissipation, are not taken into account, something that could limit the applicability of the results obtained from these studies in more realistic, open quantum systems \cite{breuer2002theory, rivas2012open}. In such systems  observe a variety of environment-induced phenomena 
 \cite{AGUDOV2003144,ghikas2012stochastic,spagnolo2012relaxation,tzemos2013dependence, magazzu2016quantum,valenti2018stabilizing,RevModPhys.95.030501, leonforte2021dressed,surazhevsky2021noise}. These phenomena play a key role in the field of quantum control \cite{cong2014control,st, stassi2016output, d2021introduction,mm}.}

{However, while the Bohmian theory has already been applied to  open quantum systems  \cite{breuer2002theory, nassar2017bohmian, oriols2017conditions, nikolic2023quantum}, no emphasis has been given yet on the role of Bohmian chaos in the evolution of open quantum systems. In fact, the field of Bohmian chaos is still under development. And at this stage, it benefits significantly from the results found by studying closed quantum systems. The insights gained from such studies provide a fundamental understanding that can be instrumental in extending the theory to more complex, open quantum systems, where nonlinear Dynamics coexists, in general, with stochastic Dynamics.}

\section{Appendix}

The theoretical average value of the $\Psi=\Psi(m_1,\sqrt{\omega_x}x)$ oscillator is
\begin{align}
\langle p_{x}\rangle=-i\hbar\int_{-\infty}^{\infty}\left(\Psi^{*}\frac{\partial\Psi}{\partial x}\right)dV,
\end{align} 
since the operator $\hat{p}_x$ is $\hat{p}_x=-i\hbar\frac{\partial}{\partial x}$. In the case of a superposition of two terms 
$\Psi=a\Psi(m_1,\sqrt{\omega_x}x)+b\Psi(m_2,\sqrt{\omega_x}x)$ we set $\hbar=1$ and write

\begin{align}
\Psi=\frac{a(\frac{\omega_x}{\pi})^{\frac{1}{4}}e^{\frac{-\omega_x x^2}{2}}H(m_1,\sqrt{\omega_x}x)e^{-it(m_1+\frac{1}{2})\omega_x}}{\sqrt{2^{m_1}m_1!}}+\frac{b(\frac{\omega_x}{\pi})^{\frac{1}{4}}e^{\frac{-\omega_x x^2}{2}}H(m_2,\sqrt{\omega_x}x)e^{-it(m_2+\frac{1}{2})\omega_x}}{\sqrt{2^{m_2}m_2!}}
\end{align}

The Hermite polynomials $H_m$ contain terms of degrees $m,m-2\dots$, i.e. of the same parity.
The conjugate wavefunction $\Psi^{*}$ is found if we set $i\to -i$ in $\Psi$ since the Hermite polynomials are real.

After some algebra we find that 
\begin{eqnarray}
-\left(\Psi^{*}\frac{\partial\Psi}{\partial x_i}\right)=\sum_i^7F_i,
\end{eqnarray}
where
\begin{eqnarray}
F_1=\frac{i\omega_x^{\frac{3}{2}}H(m_1,\sqrt{\omega_x }x)^2a^2x}{e^{\omega_x x^2}\sqrt{\pi}(a^2+b^2)2^{m_1}m_1!}+\frac{i\omega_x^{\frac{3}{2}}H(m_2,\sqrt{\omega_x }x)^2b^2x}{e^{\omega_x x^2}\sqrt{\pi}(a^2+b^2)2^{m_2}m_2!},
\end{eqnarray}

\begin{eqnarray}
F_2=\frac{i\omega_x^{\frac{3}{2}}\exp(it\omega_xm_2)H(m_2,\sqrt{\omega_x }xH(m_1,\sqrt{\omega_x }x)2^{-\frac{1}{2}(m_1+m_2)}abx}{e^{\omega_x x^2}\sqrt{\pi}\sqrt{m_1!}\sqrt{m_2!}(a^2+b^2)\exp(i\omega_xm_1t)},
\end{eqnarray}

\begin{eqnarray}
F_3=\frac{i\omega_x^{\frac{3}{2}}\exp(it\omega_xm_1)H(m_2,\sqrt{\omega_x }xH(m_1,\sqrt{\omega_x }x)2^{-\frac{1}{2}(m_1+m_2)}abx}{e^{\omega_x x^2}\sqrt{\pi}\sqrt{m_1!}\sqrt{m_2!}(a^2+b^2)\exp(i\omega_xm_2t)},
\end{eqnarray}

\begin{eqnarray}
F_4=-\frac{2i\omega_x H(m_1\sqrt{\omega_x }x)m_1H(m_1-1,\sqrt{\omega_x }x)a^2}{e^{\omega_x x^2} \sqrt{\pi}m_1!(a^2+b^2)2^{m_1}},
\end{eqnarray}

\begin{eqnarray}
F_5=-\frac{2i\omega_x H(m_2\sqrt{\omega_x }x)m_2H(m_2-1,\sqrt{\omega_x }x)b^2}{e^{\omega_x x^2} \sqrt{\pi}m_2!(a^2+b^2)2^{m_2}},
\end{eqnarray}

\begin{eqnarray}
F_6=-\frac{2i\omega_x H(m_2\sqrt{\omega_x }x)m_1H(m_1-1,\sqrt{\omega_x }x)ab 2^{-\frac{1}{2}(m_1+m_2)} \exp(i\omega_x t m_2) }{e^{\omega_x x^2} \sqrt{\pi}m_2!(a^2+b^2)2^{m_2}\exp(i\omega_x  m_1t) },
\end{eqnarray}

\begin{eqnarray}
F_7=-\frac{2i\omega_x H(m_1\sqrt{\omega_x }x)m_2H(m_2-1,\sqrt{\omega_x }x)ab 2^{-\frac{1}{2}(m_1+m_2)} \exp(i\omega_x  m_1t) }{e^{\omega_x x^2} \sqrt{\pi}m_2!(a^2+b^2)2^{m_2}\exp(i\omega_x  m_2t) }.
\end{eqnarray}
 
We note that in finding the derivative of $H_m(x)$ we used the formula $H'_m(x)=2mH_{m-1}(x)$.

The integral from $-\infty$ to $\infty$ of $F_1$ is zero since both of its terms are odd functions. Similarly both $F_4$ and $F_5$ give zero integrals since they contain different Hermite polynomials which are orthogonal.
Thus the only terms that may give a nonzero contribution in $\langle p_x\rangle$ are $F_2, F_3, F_6$ and $F_7$. The integrals of the functions $F_2$ and $F_3$ are of the form 
\begin{align}
\int_{-\infty}^{\infty}xe^{-x^2}H_m(x)H_n(x)dx=\pi^{\frac{1}{2}}2^{n-1}n!\delta_{m,n-1}+\pi^{\frac{1}{2}}2^n(n+1)!\delta_{m,n+1},
\end{align}
where $\delta_{r,s}$ stands for the Kronecker symbol (see \cite{arfken2011mathematical}).
Therefore they are nonzero only if $m, n$ differ by 1. Finally, from $F_6$ and $F_7$ we find nonzero integrals only if  $m_2=m_1-1$ or vice versa, i.e.  the $m_1$ and $m_2$ should differ by 1 otherwise they give zero integrals.

If $\Psi$ is the sum of 3 terms $\Psi=a\Psi_{m_1}+b\Psi_{m_2}+c\Psi_{m_3}$ then we have also terms similar to $F_1-F_7$ between $\Psi_{m_1}$ and $\Psi_{m_3}$ and between $\Psi_{m_2}$ and $\Psi_{m_3}$. Thus in order to have an integral different from zero we must have $|m_1-m_2|=1$ or  $|m_1-m_3|=1$ or $|m_2-m_3|=1$. If $\Psi$ depends both on $x$ and $y$, 
\begin{align}
\Psi=a\Psi_{m_1}(x)\Psi_{n_1}(y)+b\Psi_{m_2}(x)\Psi_{n_2}(y)+c\Psi_{m_3}(x)\Psi_{n_3}(y).
\end{align}

After the integration with respect to $x$ we have to integrate also with respect to $y$, as in Eq.~\ref{pxint}. The only integrals $\int\Psi_n(y)\Psi_{n'}(y)dy$ that are different from zero are those with $n=n'$. Therefore in a term with $|m_1-m_2|=1$ we must also have $n_1=n_2$, in the term $|m_1-m_3|=1$ we must also have $n_1=n_3$, and in a term $|m_2-m_3|=1$ we must also have $n_2=n_3$. 

Similar analytical expressions are found for $\langle p_y\rangle$ and $\langle L\rangle$. 
Regarding the position of the particles, the theoretical average value of $x$ in the case of two terms   gives
\begin{align}
\langle x\rangle=\int_{-\infty}^{\infty}\Psi^*x\Psi dx,
\end{align}
where $\Psi^*x\Psi$ contains terms of the form $xH_{m_1}^2(x), xH_{m_2}(x)^2,$ and $xH_{m_1}(x)H_{m_2}(x)$. The integrals of the first two terms are zero. The integral of third term is different from zero only if $m_1$ and $m_2$ differ by 1. Similar results we find when $\Psi$ contains 3 terms in $x$.

Then if $\Psi$ depends also on $y$  we must integrate also with respect to $y$. This integration gives zero unless $|m_1-m_2|=1, n_1=n_2$ or $|m_1-m_3|=1, n_1=n_3,$ or $|m_2-m_3|=1, n_2=n_3$. 

Similarly, in calculating the theoretical average value of $y$ we have a result different from zero only if $|n_1-n_2|=1$ and $m_1=m_2$ or $|m_1-m_3|=1 $ and $m_2=m_3$, or $|n_2-n_3|=1$ and $m_2=m_3$.

In the example of the single nodal point  $\Psi=M(a\Psi_{0,0}+b\Psi_{1,0}+c\Psi_{1,1})$ we have $m_1=0, m_2=1, m_3=1$ and $n_1=0, n_2=0, n_3=1$. Therefore the $x$-integration for $\langle p_x\rangle$ gives two terms different from zero, those with $|m_1-m_2|=1$ and with $|m_1-m_3|=1$. The $y$-integration of the first term gives a non-zero term because $n_1=n_2=0$ while the integration of the second term gives zero, because $n_2\neq n_3$. The final result is given by Eq.~\ref{pxms}. In a similar way we calculate $\langle p_y\rangle$ (Eq.~\ref{pyms}), $\langle L\rangle$ (Eq.~\ref{Lms}) $\langle x\rangle$ (Eq.~\ref{xms}) and $\langle y\rangle$ (Eq.~\ref{yms}).

In the example of many nodal points $\Psi=M(a\Psi_{0,2}+b\Psi_{3,4}+c\Psi_{5,7})$ (case 3.1) we have $m_1=0, m_2=3, m_3=5$ and $n_1=2, n_2=4, n_3=7$, therefore there are  no terms with successive $m_i,m_j$ and the values of $\langle p_x\rangle, \langle p_y\rangle$,  $\langle L\rangle$, $\langle x\rangle$ and $\langle y\rangle$ are zero. 

Similarly we have zero results in the case 3.2 when $\Psi=M(a\Psi_{10,3}+b\Psi_{4,5}+c\Psi_{7,8})$, while in the case $\Psi=M(a\Psi_{4,6}+b\Psi_{5,6}+c\Psi_{7,8})$ we  have $\langle p_x\rangle$ different from zero, but $\langle p_y\rangle=0$.

\section*{Acknowledgements}
This research was conducted in the framework of the program of the RCAAM
of the Academy of Athens “Study of the dynamical evolution of the entanglement and coherence in
quantum systems”.

\bibliographystyle{elsarticle-num}
\bibliography{bibliography}

\begin{thebibliography}{10}
\expandafter\ifx\csname url\endcsname\relax
  \def\url#1{\texttt{#1}}\fi
\expandafter\ifx\csname urlprefix\endcsname\relax\def\urlprefix{URL }\fi
\expandafter\ifx\csname href\endcsname\relax
  \def\href#1#2{#2} \def\path#1{#1}\fi

\bibitem{Bohm}
D.~Bohm, A suggested interpretation of the quantum theory in terms of "hidden"
  variables. i, Phys. Rev. 85 (1952) 166.

\bibitem{BohmII}
D.~Bohm, A suggested interpretation of the quantum theory in terms of "hidden"
  variables. ii, Phys. Rev. 85 (1952) 180.

\bibitem{holland1995quantum}
P.~R. Holland, The quantum theory of motion: an account of the de Broglie-Bohm
  causal interpretation of quantum mechanics, Cambridge Univ. Press, 1995.

\bibitem{bacciagaluppi2009quantum}
G.~Bacciagaluppi, A.~Valentini, Quantum theory at the crossroads: reconsidering
  the 1927 Solvay conference, Cambridge Univ. Press, 2009.

\bibitem{lazarovici2019quantum}
D.~Lazarovici, M.~Hubert, How quantum mechanics can consistently describe the
  use of itself, Sci. Rep. 9 (2019) 470.

\bibitem{valentini1991signalII}
A.~Valentini, Signal-locality, uncertainty, and the subquantum h-theorem. ii,
  Phys. Lett. A 158 (1991) 1.

\bibitem{valentini1991signalI}
A.~Valentini, Signal-locality, uncertainty, and the subquantum h-theorem. i,
  Phys. Lett. A 156 (1991) 5.

\bibitem{towler2011time}
M.~Towler, N.~Russell, A.~Valentini, Time scales for dynamical relaxation to
  the {B}orn rule, Proc. Roy. Soc. A 468 (2011) 990.

\bibitem{merzbacher1998quantum}
E.~Merzbacher, Quantum mechanics, John Wiley \& Sons, 1998.

\bibitem{shankar2012principles}
R.~Shankar, Principles of quantum mechanics, Springer, 2012.

\bibitem{ballentine2014quantum}
L.~E. Ballentine, Quantum Mechanics: a Modern Development, World Scientific,
  2014.

\bibitem{haake1991quantum}
F.~Haake, Quantum signatures of chaos, Springer, 1991.

\bibitem{Stockmann_1999}
H.-J. Stöckmann, Quantum Chaos: An Introduction, Cambridge Univ. Press, 1999.

\bibitem{wimberger2014nonlinear}
S.~Wimberger, Nonlinear dynamics and quantum chaos, Springer, 2014.

\bibitem{robnik2016fundamental}
M.~Robnik, Fundamental concepts of quantum chaos, Europ. Phys. J. Special Top.
  225 (2016) 959.

\bibitem{parmenter1995deterministic}
R.~H. Parmenter, R.~Valentine, Deterministic chaos and the causal
  interpretation of quantum mechanics, Phys. Lett. A 201 (1995) 1.

\bibitem{sengupta1996quantum}
S.~Sengupta, P.~Chattaraj, The quantum theory of motion and signatures of chaos
  in the quantum behaviour of a classically chaotic system, Phys. Lett. A 215
  (1996) 119.

\bibitem{iacomelli1996regular}
G.~Iacomelli, M.~Pettini, Regular and chaotic quantum motions, Phys. Lett. A
  212 (1996) 29.

\bibitem{frisk1997properties}
H.~Frisk, Properties of the trajectories in {B}ohmian mechanics, Phys. Lett. A
  227 (1997) 139.

\bibitem{wu1999quantum}
H.~Wu, D.~Sprung, Quantum chaos in terms of bohm trajectories, Phys. Lett. A
  261 (1999) 150.

\bibitem{makowski2000chaotic}
A.~Makowski, P.~Pep{\l}owski, S.~Dembi{\'n}ski, Chaotic causal trajectories:
  the role of the phase of stationary states, Phys. Lett. A 266 (2000) 241.

\bibitem{makowski2001simplest}
A.~J. Makowski, M.~Frackowiak, The simplest non-trivial model of chaotic causal
  dynamics, Acta Phys. Pol. B 32 (2001) 2831.

\bibitem{makowski2002forced}
A.~J. Makowski, Forced dynamical systems derivable from bohmian mechanics, Acta
  Phys. Pol. B 33 (2002) 583.

\bibitem{wisniacki2003dynamics}
D.~Wisniacki, F.~Borondo, R.~Benito, Dynamics of quantum trajectories in
  chaotic systems, Europhys. Lett. 64 (2003) 441.

\bibitem{falsaperla2003motion}
P.~Falsaperla, G.~Fonte, On the motion of a single particle near a nodal line
  in the de {B}roglie--{B}ohm interpretation of quantum mechanics, Phys. Lett.
  A 316 (2003) 382.

\bibitem{wisniacki2005motion}
D.~A. Wisniacki, E.~R. Pujals, Motion of vortices implies chaos in {B}ohmian
  mechanics, Europhys. Lett. 71 (2005) 159.

\bibitem{wisniacki2007vortex}
D.~Wisniacki, E.~Pujals, F.~Borondo, Vortex dynamics and their interactions in
  quantum trajectories, J. Phys. A 40 (2007) 14353.

\bibitem{borondo2009dynamical}
F.~Borondo, A.~Luque, J.~Villanueva, D.~A. Wisniacki, A dynamical systems
  approach to {B}ohmian trajectories in a 2d harmonic oscillator, J. Phys. A 42
  (2009) 495103.

\bibitem{cesa2016chaotic}
A.~Cesa, J.~Martin, W.~Struyve, Chaotic bohmian trajectories for stationary
  states, J. Phys. A 49 (2016) 395301.

\bibitem{santos2024broglie}
H.~Santos~Lima, M.~Paix{\~a}o, C.~Tsallis, de {B}roglie-{B}ohm analysis of a
  nonlinear membrane: From quantum to classical chaos, Chaos 34 (2024) 023125.

\bibitem{efth2009}
C.~Efthymiopoulos, C.~Kalapotharakos, G.~Contopoulos, Origin of chaos near
  critical points of quantum flow, Phys. Rev. E 79 (2009) 036203.

\bibitem{tzemos2023unstable}
A.~C. Tzemos, G.~Contopoulos, Unstable points, ergodicity and born’s rule in
  2d bohmian systems, Entropy 25 (2023) 1089.

\bibitem{tzemos2020chaos}
A.~C. Tzemos, G.~Contopoulos, Chaos and ergodicity in an entangled two-qubit
  {B}ohmian system, Phys. Scr. 95 (2020) 065225.

\bibitem{tzemos2021role}
A.~Tzemos, G.~Contopoulos, The role of chaotic and ordered trajectories in
  establishing {B}orn’s rule, Phys. Scr. 96 (2021) 065209.

\bibitem{efthymiopoulos2006chaos}
C.~Efthymiopoulos, G.~Contopoulos, Chaos in {B}ohmian quantum mechanics, J.
  Phys. A 39 (2006) 1819.

\bibitem{tzemos2020ergodicity}
A.~C. Tzemos, G.~Contopoulos, Ergodicity and {B}orn's rule in an entangled
  two-qubit {B}ohmian system, Phys. Rev. E 102 (2020) 042205.

\bibitem{horodecki2009quantum}
R.~Horodecki, P.~Horodecki, M.~Horodecki, K.~Horodecki, Quantum entanglement,
  Rev. Mod. Phys. 81 (2009) 865.

\bibitem{tzemos2019bohmian}
A.~C. Tzemos, G.~Contopoulos, C.~Efthymiopoulos, Bohmian trajectories in an
  entangled two-qubit system, Phys. Scr. 94 (2019) 105218.

\bibitem{PhysRevE.99.062209}
F.~J. Arranz, R.~M. Benito, F.~Borondo, Shannon entropy at avoided crossings in
  the quantum transition from order to chaos, Phys. Rev. E 99 (2019) 062209.

\bibitem{breuer2002theory}
H.-P. Breuer, F.~Petruccione, The theory of open quantum systems, Oxford Univ.
  Press, 2002.

\bibitem{rivas2012open}
A.~Rivas, S.~F. Huelga, Open quantum systems, Vol.~10, Springer, 2012.

\bibitem{AGUDOV2003144}
N.~V. Agudov, A.~A. Dubkov, B.~Spagnolo, Escape from a metastable state with
  fluctuating barrier, Phys. A 325 (2003) 144.

\bibitem{ghikas2012stochastic}
D.~P. Ghikas, A.~C. Tzemos, Stochastic anti-resonance in the time evolution of
  interacting qubits, Int. J. Quantum Inf. 10 (2012) 1250023.

\bibitem{spagnolo2012relaxation}
B.~Spagnolo, P.~Caldara, A.~La~Cognata, G.~Augello, D.~Valenti, A.~Fiasconaro,
  A.~Dubkov, G.~Falci, Relaxation phenomena in classical and quantum systems,
  Acta Phys. Pol. B 43 (2012) 1169.

\bibitem{tzemos2013dependence}
A.~C. Tzemos, D.~P. Ghikas, Dependence of noise induced effects on state
  preparation in multiqubit systems, Phys. Lett. A 377 (2013) 2307.

\bibitem{magazzu2016quantum}
L.~Magazz{\`u}, A.~Carollo, B.~Spagnolo, D.~Valenti, Quantum dissipative
  dynamics of a bistable system in the sub-ohmic to super-ohmic regime, J.
  Stat. Mech. 2016 (2016) 054016.

\bibitem{valenti2018stabilizing}
D.~Valenti, A.~Carollo, B.~Spagnolo, Stabilizing effect of driving and
  dissipation on quantum metastable states, Phys. Rev. A 97 (2018) 042109.

\bibitem{RevModPhys.95.030501}
G.~Parisi, Nobel lecture: Multiple equilibria, Rev. Mod. Phys. 95 (2023)
  030501.

\bibitem{leonforte2021dressed}
L.~Leonforte, D.~Valenti, B.~Spagnolo, A.~Carollo, F.~Ciccarello, Dressed
  emitters as impurities, Nanophotonics 10 (2021) 4251.

\bibitem{surazhevsky2021noise}
I.~Surazhevsky, V.~Demin, A.~Ilyasov, A.~Emelyanov, K.~Nikiruy, V.~Rylkov,
  S.~Shchanikov, I.~Bordanov, S.~Gerasimova, D.~Guseinov, et~al.,
  Noise-assisted persistence and recovery of memory state in a memristive
  spiking neuromorphic network, Chaos, Solit. Fractals 146 (2021) 110890.

\bibitem{cong2014control}
S.~Cong, Control of quantum systems: theory and methods, John Wiley \& Sons,
  2014.

\bibitem{st}
R.~Stassi, S.~De~Liberato, L.~Garziano, B.~Spagnolo, S.~Savasta, Quantum
  control and long-range quantum correlations in dynamical casimir arrays,
  Phys. Rev. A 92 (2015) 013830.

\bibitem{stassi2016output}
R.~Stassi, S.~Savasta, L.~Garziano, B.~Spagnolo, F.~Nori, Output
  field-quadrature measurements and squeezing in ultrastrong cavity-qed, New J.
  Phys. 18 (2016) 123005.

\bibitem{d2021introduction}
D.~d’Alessandro, Introduction to quantum control and dynamics, CRC, 2021.

\bibitem{mm}
G.~D. Fresco, B.~Spagnolo, D.~Valenti, A.~Carollo, {Multiparameter quantum
  critical metrology}, SciPost Phys. 13 (2022) 077.

\bibitem{nassar2017bohmian}
A.~B. Nassar, S.~Miret-Art{\'e}s, Bohmian {M}echanics, open quantum systems and
  continuous measurements, Springer, 2017.

\bibitem{oriols2017conditions}
X.~Oriols, A.~Benseny, Conditions for the classicality of the center of mass of
  many-particle quantum states, New J. Phys. 19~(6) (2017) 063031.

\bibitem{nikolic2023quantum}
H.~Nikolic, Quantum statistical mechanics from a {B}ohmian perspective,
  arXiv:2308.10500.

\bibitem{arfken2011mathematical}
G.~B. Arfken, H.~J. Weber, F.~E. Harris, Mathematical methods for physicists: a
  comprehensive guide, Academic press, 2011.

\end{thebibliography}

\end{document}